\documentclass{aa}
\usepackage{graphicx,natbib,amssymb,txfonts}
\bibpunct{(}{)}{;}{a}{}{,}

\renewcommand{\deg}{^{\circ}}

\begin{document}

\title{Iron lines from transient and persisting hot spots\\on AGN accretion disks}

\author{R.~W. Goosmann\inst{1,2},
        M.~Mouchet\inst{2,3},
        B.~Czerny\inst{2,4},
        M.~Dov\v{c}iak\inst{1},
        V.~Karas\inst{1},
        A.~R\'o\.za\'nska\inst{2,4},
        \and
        A.-M.~Dumont\inst{2}}

\institute{
$^1$~Astronomical Institute, Academy of Sciences, Bo{\v c}n\'{\i}~II~1401,
     CZ--14131~Prague, Czech~Republic\\
$^2$~Observatoire de Paris, Section de Meudon, LUTH, 5 place Jules Janssen,
     F--92195~Meudon Cedex, France\\
$^3$~Laboratoire Astroparticule et Cosmologie, Universit\'e Paris~7,
     10~rue Alice Domon et L\'eonie Duquet, F--75205~Paris Cedex 13, France\\
$^4$~Copernicus Astronomical Center, Bartycka 18, P--00\,716~Warsaw, Poland}

\offprints{R.~W. Goosmann\\ email:goosmann@astro.cas.cz}

\authorrunning{Goosmann et al.}

\titlerunning{Iron lines from transient versus persistent hot spots}

\abstract{}
{We model the X-ray reprocessing from a strong co-rotating flare above an
  accretion disk in active galactic nuclei. By performing detailed radiative
  transfer computations we explore the horizontal structure and evolution of
  the underlying hot spot. The main goal is to study how the resulting spectral
  features manifest themselves in short exposure time spectra.}
{We analyze both the vertical and the horizontal spot structure and its
  dynamical reprocessed spectrum. To obtain the spectral evolution seen by a
  distant observer, we apply a general relativity ray-tracing technique. We
  concentrate on the energy band around the iron K-line, where the
  relativistic effects are most pronounced. Persistent flares lasting for a
  significant fraction of the orbital time scale and short, transient flares
  are considered.}
{In our time-resolved analysis, the spectra recorded by a distant observer
  depend on the position of the flare/spot with respect to the central black
  hole. If the flare duration significantly exceeds the light travel
  time across the spot, then the spot horizontal stratification is
  unimportant. On the other hand, if the flare duration is comparable to the
  light travel time across the spot radius, the lightcurves exhibit a typical
  asymmetry in their time profiles. The sequence of dynamical spectra proceeds
  from more strongly to less strongly ionized re-emission. At all locations
  within the spot the spectral intensity increases towards edge-on emission
  angles, revealing the limb brightening effect.}
{Future X-ray observatories with significantly larger effective collecting
  areas will enable to spectroscopically map out the azimuthal irradiation structure
  of the accretion disk and to localize persistent flares. If the hot spot is
  not located too close to the marginally stable orbit of the black hole, it
  will be possible to probe the reflecting medium via the sub-structure of the iron
  K-line. Indications for transient flares will only be obtained from analyzing
  the observed lightcurves on the gravitational time scale of the accreting
  supermassive black hole.}


\date{Received ...; accepted ...}

\maketitle

\section{Introduction}

The broad shape of the iron line complex seen in about 25\% of all active
galactic nuclei (AGN) is commonly assumed to be due to relativistic effects
acting in the immediate vicinity of the central, supermassive black hole
\citep[e.g.][]{tanaka1995, iwasawa1999, wilms2001, lubinski2001, fabian2002,
young2005, guainazzi2006}. The line is expected to be a reprocessing feature
emitted by the surface of the irradiated inner accretion disk. An attractive
scenario for the origin of the X-ray irradiation envisions the reconnection of
magnetic loops above the disk surface (see e.g. Galeev, Rosner \& Vayana 1979;
Abramowicz et al. 1991; Haardt et al. 1994; Collin et al. 2003, Merloni \&
Fabian 2001). The line emitting region lies at distances of only a few
gravitational radii $R_{\rm g} = \frac{GM}{c^2}$ from the disk center.

Recently, some spectrally resolved variability of the line complex was
observed for several objects \citep[i.e.][]{iwasawa2004, turner2006, miller2006,
tombesi2007, miniutti2007}. For two of them, NGC~3516 and NGC~3783, the iron
line feature appears to vary systematically in flux and centroid energy. The
variations happen on characteristic time-scales of 25 ks and 27 ks
respectively \citep{iwasawa2004, tombesi2007}. \Citet{iwasawa2004} suggest a
promising explanation for this type of line behavior invoking an X-ray flare,
which is co-rotating with the accretion disk. The flare irradiates the disk
surface creating a hot spot. In the case of NGC~3516 the flare and the
underlying hot spot are located at (7--16)~$R_{\rm g}$  from
the black hole. The systematic changes of the reprocessed spectrum are then
explained by the orbital motion of the disk and the resulting time-dependent
relativistic modifications.

The above mentioned model requires a prescription for the reprocessed emission
from the disk. The modeling of the reprocessed spectra is often carried out
with considerable simplifications. Many computations assume either a uniform
density of the disk atmosphere, or semi-isotropic illumination and
``observation''. The angular dependence of the reflected component formed by a
constant density partially ionized medium was studied starting with Matt,
Fabian \& Ross (1993a,b), \. Zycki et al. (1994), \. Zycki \& Czerny
(1994). Then, reprocessing from a medium in hydrostatic equilibrium was
modeled by Raymond (1993), Nayakshin et al. (2000), Ballantyne et al.  (2001),
Ballantyne \& Ross (2002), R\'o\.za\'nska et al. (2002).  Nayakshin \& Kazanas
(2002) pioneered the simplified time-dependent reflection studies for
photoionized accretion disks. \citet{goosmann2006a} studied rms spectral
variability from a distribution of flares with vertically stratified spots
underneath.  None of those models considered a horizontal stratification of the
illuminated spot as suggested by \citet{goosmann2006b}. 

In this paper we add considerable sophistication to the flare model by
studying in detail the Compton reflection/reprocessed component coming from a
single orbiting spot. Since the spot properties have a significant
gradient from the spot center to the border, we take both the vertical
stratification of the reprocessing material and the horizontal stratification
into account. The locally re-emitted spectrum is computed at various emission
angles, and combined with a relativistic ray-tracing method
\citep{dovciak2004a,dovciak2004b} to model the spectra seen by a distant observer.

We model the observed time evolution of the hot spot emission for two cases: a
persisting flare lasting for a significant fraction of one orbital period at
the given disk radius and short-term flares with durations of a few hundreds
of seconds. The first case applies to recently reported observations of
exceptionally powerful flares lasting for minutes to hours. Examples were given
e.g. for MCG-6-30-15 \citep[see Fig.~7 in][]{ponti2004} and NGC~5548
\citep[see Fig.~1 in][]{kaastra2004}.  For the short-term flares, the horizontal
stratification of the spot becomes important. During the onset and fading of the
flare, the spot evolves from the center, where the first primary photons
arrive, to the border. We calculate this effect presenting the resulting
lightcurves and averaged spectra for short-term flares. We show that future data
from the planned {\it XEUS} satellite will be of enough quality to be compared
with our models.

The structure of the paper is as follows: section~\ref{sect:model} describes
our model. The obtained locally emitted spectra are presented in
Sect.~\ref{sec:local-prop}, and those seen by a distant observer in
Sect.~\ref{sec:distant-spec}. The importance of the studied effects is
discussed in Sect.~\ref{sec:effects}. Broader implications and some final
conclusions are given in Sect.~\ref{sec:discuss}.

\section{The model}
\label{sect:model}

Our method to compute the local spectra of the hot spot is similar to the one
described in \citet{collin2003}. First, we compute the vertical density
structure of the disk at a given radius $R$ assuming that the disk is in
hydrostatic equilibrium without any external irradiation. The density
structure of the surface layers are then used to conduct detailed radiative
transfer computations. We take into account the intrinsic time evolution of a
flare and we compute the spectral changes as seen by a distant observer. We
apply a ray-tracing technique to include appropriate time-delays, general
relativity effects, and the orbital motion of the hot spot.

In this paper, we assume for all computations the metric of a Schwarzschild
black hole with $M=10^8 M_\odot$ and an accretion rate $\dot m=0.001$ in units
of the Eddington accretion rate. The surrounding disk with the viscosity
parameter $\alpha=0.1$ is considered at the radii 7~$R_{\rm g}$ and 18~$R_{\rm
  g}$.

\subsection{The vertical disk structure}
\label{sec:hydro-equi}

Following the argument made by Collin et al. (2003) we assume that the
flare is much shorter than the characteristic timescale for the
restoration of the hydrostatic equilibrium. Hence, we assume that there
is no external irradiation before the onset of the flare and the
computation of the initial disk profiles is based only on the emission
coming from the standard $\alpha$-disk at the given radius $R$. This
condition is safely fulfilled if we assume that the flare duration is less
than the orbital timescale at $R$.

We assume that the disk atmosphere is a plane-parallel, vertically stratified
slab. The density, temperature, and radiation flux profiles are computed using
an advanced version of the code described in  \citet{rozanska1996},
\citet{rozanska1999}, and R\'o\.za\'nska et al. (2002).  The code computes the
hydrostatic balance of the medium in the gravitational field of the black
hole, thereby including the relativistic corrections to the equations of the
vertical disk structure \citep{novikov1973, page1974}.  The radiation inside
the medium is transported by diffusion using the Rossland approximation.

\subsection{Computing the local spectra}
\label{sec:geom-aspects}

Having obtained the vertical density profile of the disk, we compute the local
reprocessed spectra expected after the onset of the flare. Since the principal
spectral features are produced within a few Thomson optical depths $\tau_{\rm
ES}$ from the disk surface, we only consider the upper layers up to $\tau_{\rm
ES} = 7$ for the radiative transfer calculations. We call this upper layer the
disk atmosphere and assume that it is heated from underneath by the thermal
radiation of the underlying disk bulk. The temperature and incident flux
$F_{\rm disk}$ of the disk bulk are computed when obtaining the vertical disk
structure. From above, the disk atmosphere is irradiated by the flare located
at a given height $h$ that is measured from the disk surface where the flare
causes the incident flux $F_{\rm inc}$.

We assume that half of the primary radiation reaches the disk atmosphere. The
main fraction of the flux is reprocessed directly underneath the flare
source. The radiation emitted further sideways reaches the disk farther away
from the flare and so it becomes geometrically diluted. Therefore, we can
define a limited hot spot on the disk surface. The spot is centered on the
orthogonal projection of the point-like flare source. The circumference of the
spot and the flare source define an irradiation cone with half-opening angle
$\theta_0$. We neglect any irradiation outside this cone, thus we only
consider reprocessed radiation coming from inside the spot.

The distance $R$ of the flare from the central black hole is measured for the
spot center. We limit the flare cone to $\theta_0 = 60 \deg$ and the height of
the source to $h = 0.5 R_{\rm g}$. We thus assume that the spot size is small
and neglect the change in the disk's internal properties across the spot. It
remains to be seen whether such an assumption is justified. The MHD simulations
of Miller \& Stone (2000) and Blaes, Hirose \& Krolik (2007) suggest magnetic
loop heights that are by $\sim 2-3$ times higher than the disk thickness,
while Turner (2004) found little evidence for magnetic loops rising so high.

\begin{figure}
  \centering
  \includegraphics[angle=90,width=\columnwidth]{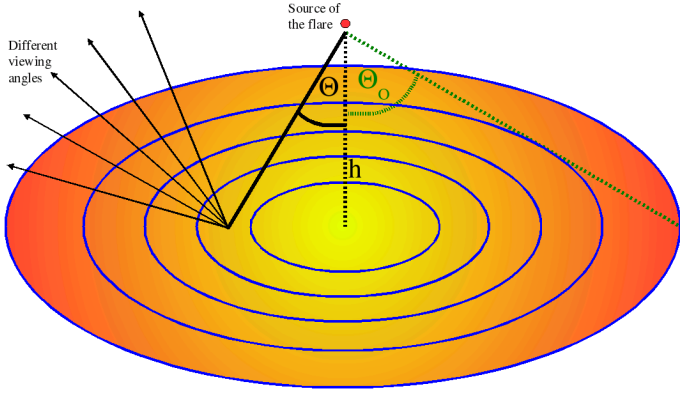}
  \caption{Illustration of the hot spot at the disk surface underlying the
  flare. The illuminated region is divided into concentric rings of equally
  intercepted fractions of the flare luminosity.}
  \label{fig:flare-geom}
\end{figure}

The total incident flux, $F_{\rm inc}^{\rm tot}$, of the whole spot is
parameterized by the ratio between the total incident flux and the internal
flux emitted by the disk, $F_{\rm disk}$. We set $F_{\rm inc}^{\rm tot}/F_{\rm
disk} = 144$. This ratio ensures that the incident radiation by the flare is
much stronger than the disk emission (the same value was used earlier by
Ballantyne et al. 2001 and Collin et al. 2003). The local incident flux,
$F_{\rm inc}$, decreases with the distance from the spot center and the
incident angle, $\theta_{\rm i}$ increases as illustrated in
Fig.~\ref{fig:flare-geom}.  We divide the hot spot into five concentric rings
defining equally spaced solid angles with respect to the flare
source. Integrated over its surface each ring thus receives the same energy per
unit time. We solve the radiative transfer for each ring individually using
separate values of $F_{\rm inc}$ and $\theta_{\rm i}$. Across a given ring we
assume that $F_{\rm inc}$ and $\theta_{\rm i}$ are constant.

The spectral shape of the primary radiation is modeled as a power law
extending from 1~eV to 100~keV  with the photon index $\Gamma = 1.9$.  The
radiative transfer  simulations are then conducted by the coupled use of the
codes {\sc Titan} and {\sc Noar} described by \citet{dumont2000} and updated
in \citet{dumont2003}. The code {\sc Titan} calculates the vertical ionization
and temperature structure of the X-ray illuminated gas and the reprocessed
spectrum in the optical, UV, and X-ray ranges. The radiative transfer equation
is solved in multi-stream approximation by the Accelerated Lambda iteration
(ALI) method for the continuum and for the lines. The resulting ionization and
temperature data are then read by the Monte-Carlo code {\sc Noar}. It
calculates heating and cooling rates due to Compton scattering in the medium
and the final emerging X-ray spectra for the reprocessed radiation in the
range of (0.8--100)~keV. The resulting spectra are computed at 20
different emission angles $\psi$ measured with respect to the disk normal
and varying in constant steps of $\cos{\psi}$.

\subsection{Keplerian motion and relativistic effects}
\label{sec:rel-effects}

A distant observer detects an evolving spectrum to which all parts of the
orbiting hot spot contribute. This contribution depends on the local spectral
emissivity but also on the relativistic effects and on the Doppler shift
acting at a given position.

The radiation propagation for a Keplerian accretion disk around a black hole
is computed by the code {\sc KY} \citep{dovciak2004a,dovciak2004b}. In our
application, {\sc KY} integrates the photon geodesics between a position
inside the orbiting hot spot and an observer at infinity. The integration
method takes into account the energy shift of the photon along the geodesics,
arrival time-lags, and the lensing-effect. Spectra are computed for various
viewing directions, $i$, of a distant observer based on a given time-dependent
emissivity distribution of the flare spot. We define the initial position of
the spot center in polar coordinates $R_0$ and $\phi_0$. With $h = 0.5 \, R_{\rm
g}$ and $\theta_0 = 60\deg$, the spot radius, $r_0$, follows to be $r_0 =
\sqrt{3} h = 0.866 \, R_{\rm g}$.

The flare above the disk surface is assumed to co-rotate with the underlying
disk. This is a natural assumption if the flares form around reconnecting
magnetic field lines that are anchored in the disk interior. The spot center
follows the same Keplerian orbit. Since the spot originates from external
illumination, we can assume that its initial shape remains circular. Nevertheless, in
principle the kinematics of the spot is affected by the differential rotation
of the disk: spot regions located closer to the black hole have a larger
angular velocity than spot regions farther away. Therefore there is a relative
drift of the material across the spot, with some material leaving the spot
region and cooling off while new material enters and is heated. We assume that
at each part of the spot the orbital motion is Keplerian but, for simplicity,
we neglect the relative drift of the material and preserve the spot's circular shape
We thus suppose that the heating and cooling of the drifting material is
instantaneous. This is a safe assumption following the estimates by Collin et
al. (2003).

In this paper, we consider the relativistic light bending only for the
radiation being re-emitted by the hot spot. Modeling the relativistic effects
for the incident radiation is not necessary because the flare occurs at a
small height. We checked this approximation conducting ray-tracing
computations for incident photons being emitted along the limiting cone of the
flare at $\theta_0=60\deg$. In flat space, these photons trace out a circle on
the surface plane of the disk. For a flare in Schwarzschild metric at radius
7~$R_{\rm g}$ and height $0.5 \, R_{\rm g}$, the relativistic trace does not
differ significantly from the flat-space result.

\subsection{Time evolution of flare and spot}
\label{sec:parameters}

The time evolution of the flare is based on two aspects. First, there is an
intrinsic time evolution of the flare source, which is parameterized by a
specific profile. For simplicity we assume that the intrinsic flare emission
has a rectangular shape of length $T$. The source is thus switched on and off
instantaneously. Second, there is a delayed response from the various parts of
the spot, as the primary radiation emitted by the elevated source reaches each
concentric ring of the hot spot consecutively. During onset and fading of the
flare, the hot spot evolves from the center, where the first primary photons
arrive, to the border. This evolution adds time-dependent effects to the
spectrum that are different from those caused by relativity and by the orbital
spot motion. Let the irradiation at the spot center $(R_0, \phi_0)$ start at
time $t_0$ and end there at $t_0+T$. Then the local emissivity $\varepsilon$
at the photon energy $E$, given disk position $(R,\phi)$, emission angle
$\psi$, and time $t$ is determined by:

\begin{equation}
  \varepsilon(E,R,\phi,\psi,t) =
  \left \{
  \begin{array}{cl}
    \varepsilon_{\rm spot}(E,R,\phi,\psi), & t_0 < t - \eta < t_0+T,\\
    0,                                     & {\rm otherwise},
  \end{array}
  \right.
\end{equation}

\noindent with $\eta = (\sqrt{h^2+r^2}-h)/c$, and the spatial distance between
the spot center and the point $(R,\phi)$ is measured by $r = \sqrt{R^2 + R_0^2
  - 2 R R_0 \cos(\phi - \phi_0)}$. The local spot emission as computed from
the radiative transfer simulations is denoted by $\varepsilon_{\rm spot}$. If
the duration of the flare is much longer than the light travel time across the
spot, the irradiation can be considered as a stationary process and the onset
and fading phases are negligible. Such a time-steady hot spot is described by
the local disk emissivity 

\begin{equation}
  \varepsilon(E,R,\phi,\psi) =
  \left \{
  \begin{array}{cl}
    \varepsilon_{\rm spot}(E,R,\phi,\psi), & r < r_0,\\
                                        0, & \rm{otherwise}. 
  \end{array}
  \right.
\label{eq:steady}
\end{equation}

Comparison with the observational data requires relativistic ray-tracing and
time integration of the local spot emissivity while taking into account the
orbital motion. These computations are consistently carried out by {\sc KY}.

\section{The local properties of the reprocessed spectra}
\label{sec:local-prop}

The overall response of the irradiated disk to the incident flux is in
agreement with known results: a hot layer forms in the upper disk atmosphere,
being at a temperature comparable to the Compton temperature (Raymond 1993, Ko
\& Kallman 1994, Nayakshin et al. 2000, Ballantyne et al. 2001, R\'o\.za\'nska
et al. 2002). At larger optical depth a rapid transition to colder, less
ionized layers follows. Nevertheless, as we show in the following, the
horizontal stratification of the hot spot we introduced in our model brings
new aspects to the structure of the medium and to the obtained reprocessed
spectra.

\subsection{Temperature profiles and reprocessed spectra}

\begin{figure*}
\centering
  \includegraphics[width=0.985\textwidth]{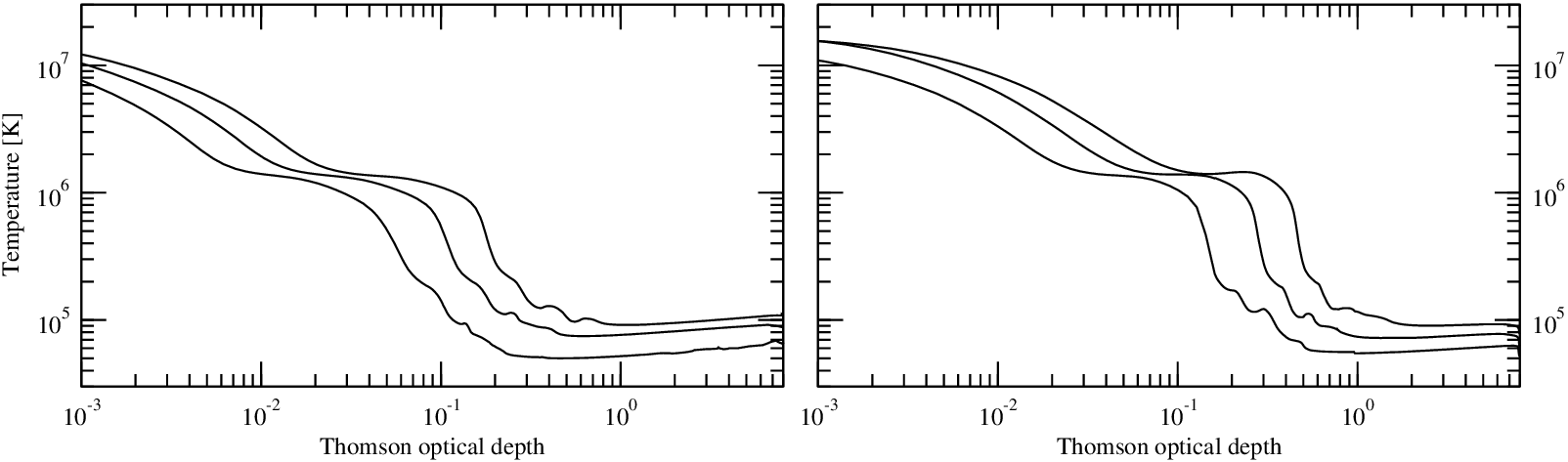}
  \vskip 0.25cm
  \includegraphics[width=\textwidth]{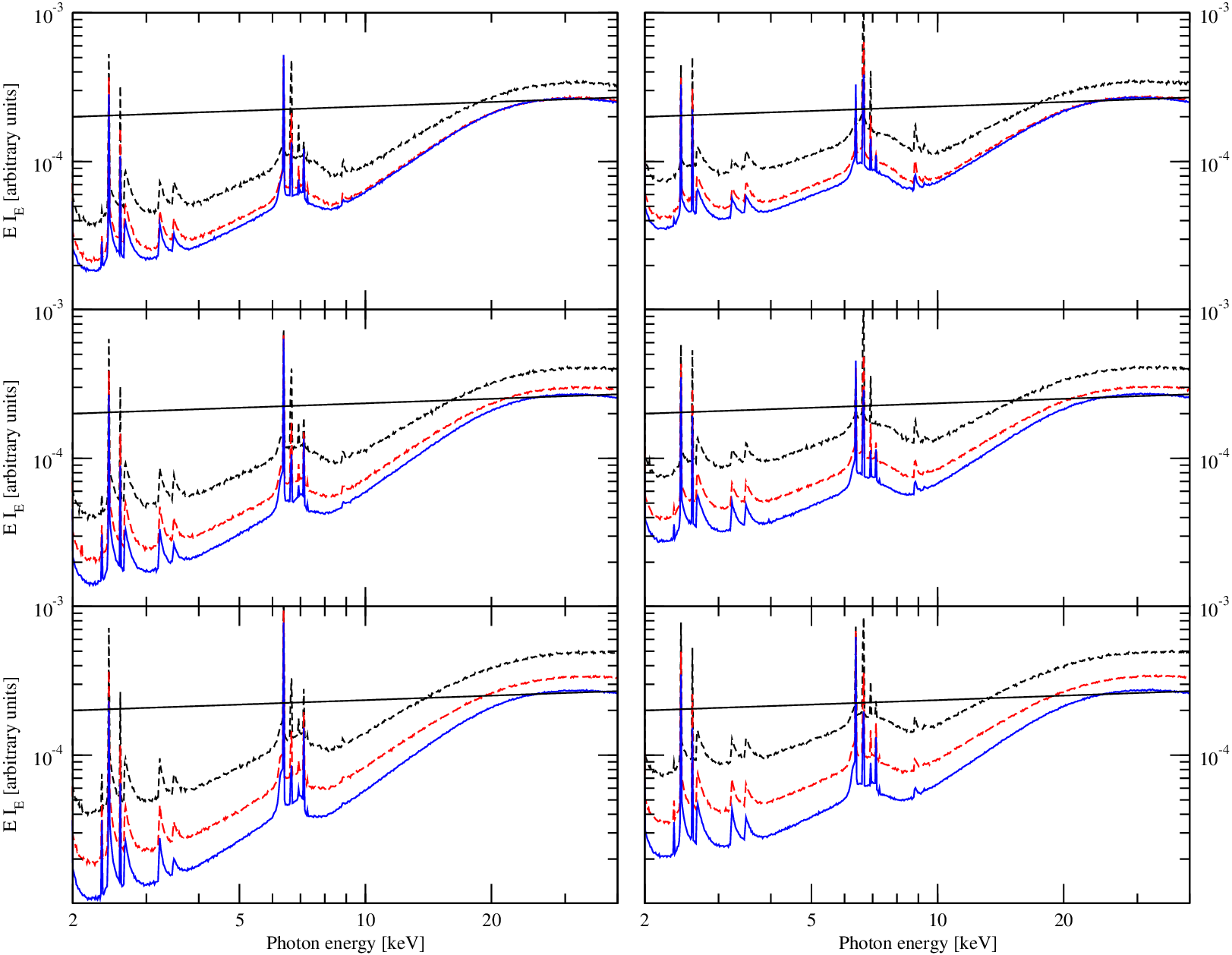}
  \caption{The top graph shows the vertical temperature profiles for three
    concentric rings of the hot spot. They correspond to incident angles of
    $\theta = 18\deg$ (inner ring, top curves), $\theta = 41\deg$ (intermediate
    ring, middle curves), and $\theta = 57\deg$ (outer ring, bottom curves). In
    the same order, the three panels of the bottom graph show the corresponding
    local reprocessed intensity spectra in $E \times I_{\rm E}$ at three different
    emission angles: $\psi = 20\deg$ (blue, solid), $\psi = 60\deg$ (red, long
    dashes), and $\psi = 80\deg$ (black, short dashes). The primary spectrum is
    indicated by the straight lines. The angles $\theta_0$ and $\psi$ are measured
    with respect to the disk normal. Results for a hot spot at $R = 7 \, R_{\rm
    g}$ (left) and $R = 18 \, R_{\rm g}$ (right) are presented.
    \label{fig:r3-i4-Ballan-spect}}
\end{figure*}

Figure~\ref{fig:r3-i4-Ballan-spect} (top graph) shows the vertical temperature
structure of the disk after the onset of the flare. We plot the temperature as
a function of the Thomson optical depth and for the medium underneath three of
the five concentric spot rings (inner, intermediate, and outer region of the
spot). The profiles are plotted versus the Thomson optical depth $\tau_{\rm
ES}$, which increases toward the equatorial plane of the disk. The hot surface
layer of the disk atmosphere and the abrupt transition to the colder medium
are visible. At the spot center the temperatures are higher, the hot skin is
thicker, and the less ionized layers lie deeper in the medium. For a given
spot ring, the heated layer at 7~$R_{\rm g}$ is narrower than at 18~$R_{\rm
g}$ which is determined by the different density profiles as computed before
irradiation by the flare.

The local reprocessed spectra emitted by different spot rings are shown in
Fig.~\ref{fig:r3-i4-Ballan-spect} (bottom graph). Going from top to bottom,
the three panels present results for the inner, intermediate, and outer spot
ring respectively. The three intensity spectra of each panel correspond to
emission angles of $\psi=20\deg$ (solid line), $\psi=60\deg$ (long-dashed
line), and $\psi=80\deg$ (short-dashed line). At the lowest $\psi$,
corresponding to a Seyfert-1 view, the Compton hump over (10--60)~keV
appears similar for all rings. At higher $\psi$, the normalization of the
emitted spectra increases and the spectral slope softens from the spot center
toward the border. For a given ring, the spectral normalization rises with
$\psi$ constituting a \emph{limb-brightening effect} of the hot spot. Such
behavior had been first seen in models of irradiated constant density media
\citep[Fig.~7 of][]{zycki1994b}.

The limb-brightening can be partly understood from the vertical temperature
structure of the medium: The upper layers of the disk surface are very hot,
therefore highly ionized, and optically thin. A local observer looking at the
medium at a normal direction $\psi \sim 0\deg$ could see down to the deeper,
less ionized layers, because along this line-of-sight the temperature drops
significantly within $\tau_{\rm ES} = 1$. However, a local observer looking at
the same medium from a higher inclination $\psi \sim 60\deg$ would rather see
the highly ionized surface layers dominated by electron scattering and
exhibiting less absorption. Along this line-of-sight the extinction within
$\tau_{\rm ES} = 1$ is dominated by scattering and thus more intensity is
preserved than for the observer at low inclinations. 

\begin{figure*}
  \centering
  \includegraphics[width=8.3cm]{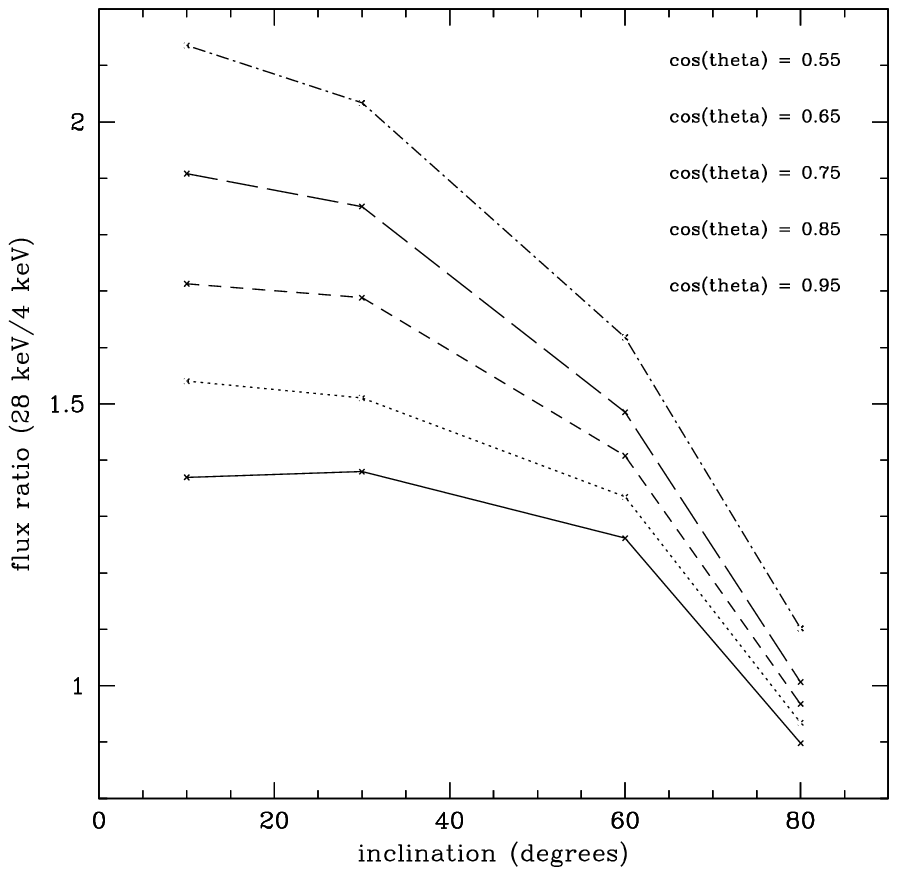}
  \hfill
  \includegraphics[width=8.3cm]{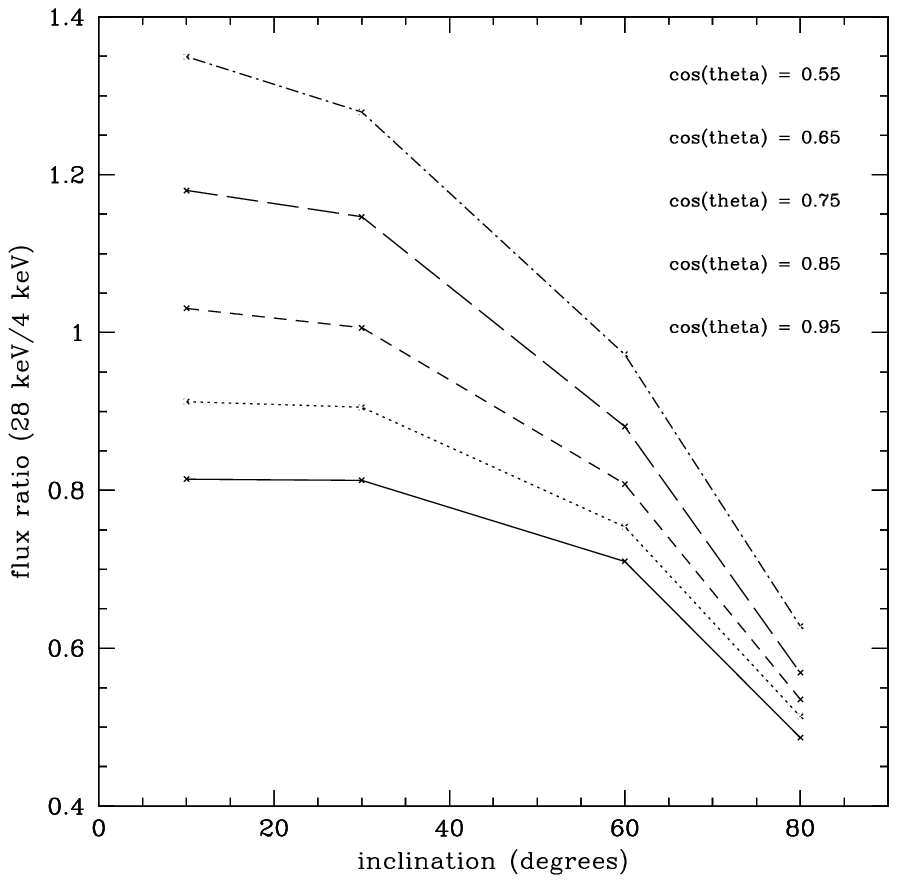}
  \caption{Spectral intensity ratios between 28~keV and 4~keV plotted versus
  the local emission angle $\psi$ (local disk inclination) for a hot spot at
  $7 R_{\rm g}$ (left) and at 18~$R_{\rm g}$ (right). The five curves denote
  the consecutive concentric rings of the hot spot. They are parameterized by
  the cosine of the incident angle $\theta$ of the illuminating primary
  radiation. The angles $\theta$ and $\psi$ are measured with respect to the
  disk normal. \label{fig:r5-slopes}}
\end{figure*}

We also investigate the dependencies of the reprocessed continuum on $\psi$ by
considering spectral intensity ratios over narrow energy bands at 4~keV
and at 28~keV. The results are plotted in Fig.~\ref{fig:r5-slopes}. 
The plot shows that the spectra become harder toward
lower emission angles. This behavior varies quantitatively with the location
inside the spot. At the spot center the irradiating
flux is strongest and the softest spectra appear. This is true for all
possible emission angles and can be understood by the relatively thicker
high-temperature layer at the disk surface. It prevents a large fraction of
the incident photons from reaching the deeper, colder medium, where they can
be absorbed. Their chance to escape from the atmosphere after a few
electron scattering events is higher than at the spot border, where the medium
absorbs more efficiently.

\subsection{The iron fluorescence complex}

The spectral appearance of the iron K-line complex also varies with $\psi$ and
with the position inside the hot spot. Generally, several components of the
line complex are visible: one at 6.4~keV representing weakly ionized/neutral
iron, the component of helium-like iron at 6.7~keV, and sometimes the
hydrogen-like iron at 6.9~keV. A K$\beta$-line at 7.1~keV is visible when the
weakly ionized component at 6.4~keV is strong.  For a given spot ring, the
lines emitted at high $\psi$ indicate stronger ionization than at low
$\psi$. The effect is analogous to the limb-brightening discussed above and
can be again understood from the temperature and ionization structure of the
medium.

The iron K-lines and also the soft X-ray emission lines of the local spectra
we compute are particularly strong with respect to the continuum. This is
partly explained by the fact that we only consider the pure reprocessed
radiation and we leave aside any dilution by the primary component. Another
reason is that in our model the relatively cold disk matter at roughly
$10^5$~K is irradiated by hard X-rays corresponding to a temperature of
($10^7$--$10^8$)~K. Since we assume that the flare duration is significantly
shorter than the dynamical timescale of the medium, the density structure of
the disk atmosphere does not change. For longer flare durations, the medium
must thermally expand, as shown in \citet{czerny2004a} and R\'o\.za\'nska et
al. (2002). During the expansion the density decreases leading to weaker line
emission \citep{collin2003}. This effect can be important for flares
completing more than a single orbit, as the one analyzed in NGC~3516 by
\citet{iwasawa2004}.

\section{Spectra seen by a distant observer}
\label{sec:distant-spec}

To obtain the reprocessed spectra seen by a distant observer, we investigate
the effects of the intrinsic time evolution of a flare combined with  general
relativity and with the orbital motion in the vicinity of the black hole. We
apply the ray-tracing code {\sc KY} with the local single-ring spectra
described in Sect.~\ref{sec:local-prop} to define the emissivity of the spot.
Note that in our modeling we focus on the re-emission of the spot and we do
not take into account the primary emission. We thus assume that the flare
source is shielded from the observer's direct view and that it manifests itself
mainly by its reflection.

\subsection{Persistent spots at different orbital phases}
\label{sec:SM-whole-orbit}

In a first step, we consider the case when the flare duration is much longer 
than the light crossing time of the hot spot. We thus neglect the onset and
the decay phase of the flare when the spot is only partly illuminated. In the
following we refer to this type of flares as {\it persistent} flares. To
define the local emission we use the time-steady spot approximation (see
Eq.~\ref{eq:steady}). Beside the location of the flare $R_0$, $\phi_0$, $h$,
and the size of the spot $\theta_0$, $r_0$ given in
Sect.~\ref{sec:rel-effects}, we assume a flare duration equal to 1/8 of the
local orbital period.

\subsubsection{Spectral appearance}

\begin{figure*}
  \centering
  \includegraphics[width=8.4cm]{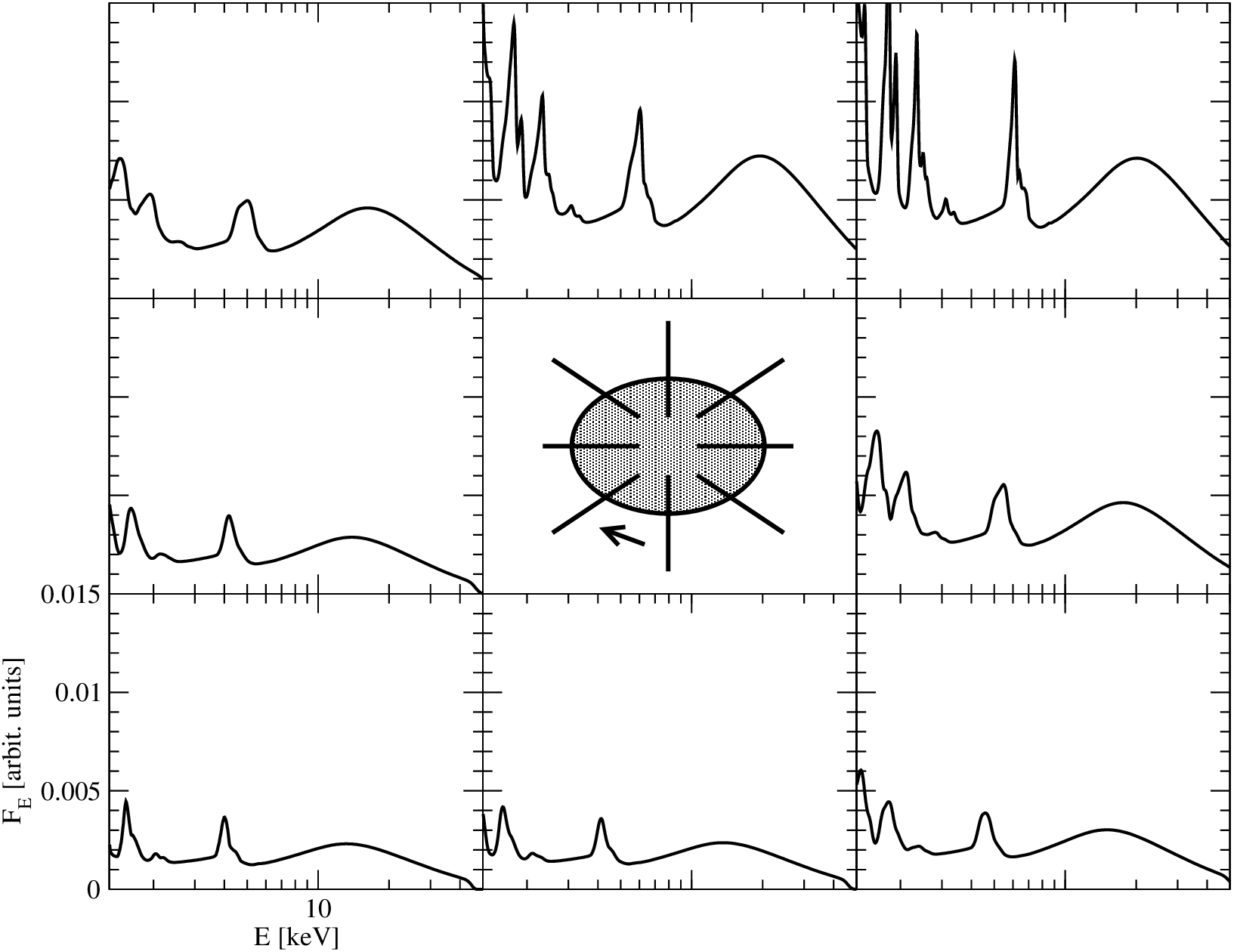}
  \hfill~
  \includegraphics[width=8.4cm]{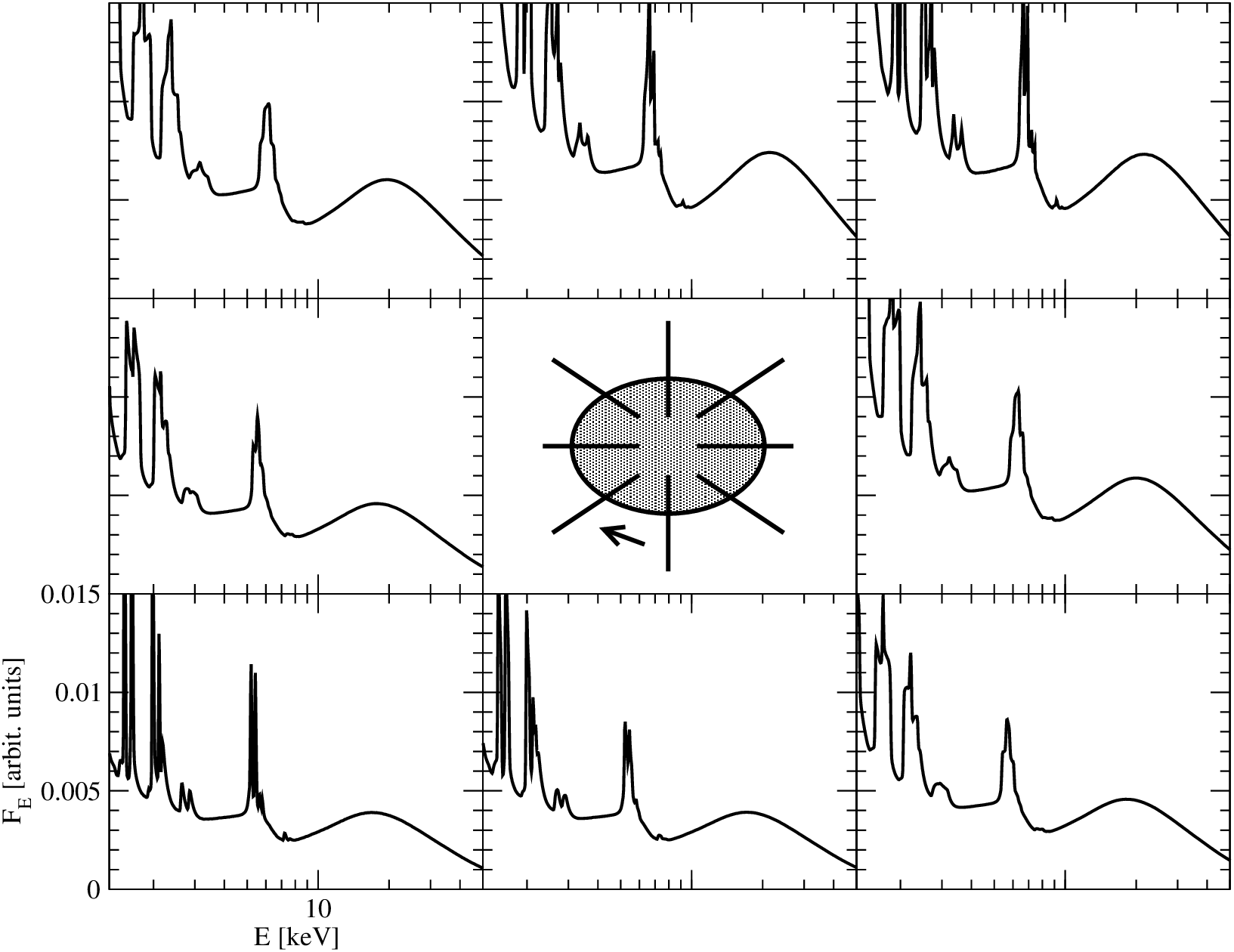}
  \vskip 0.3 truecm
  \includegraphics[width=8.4cm]{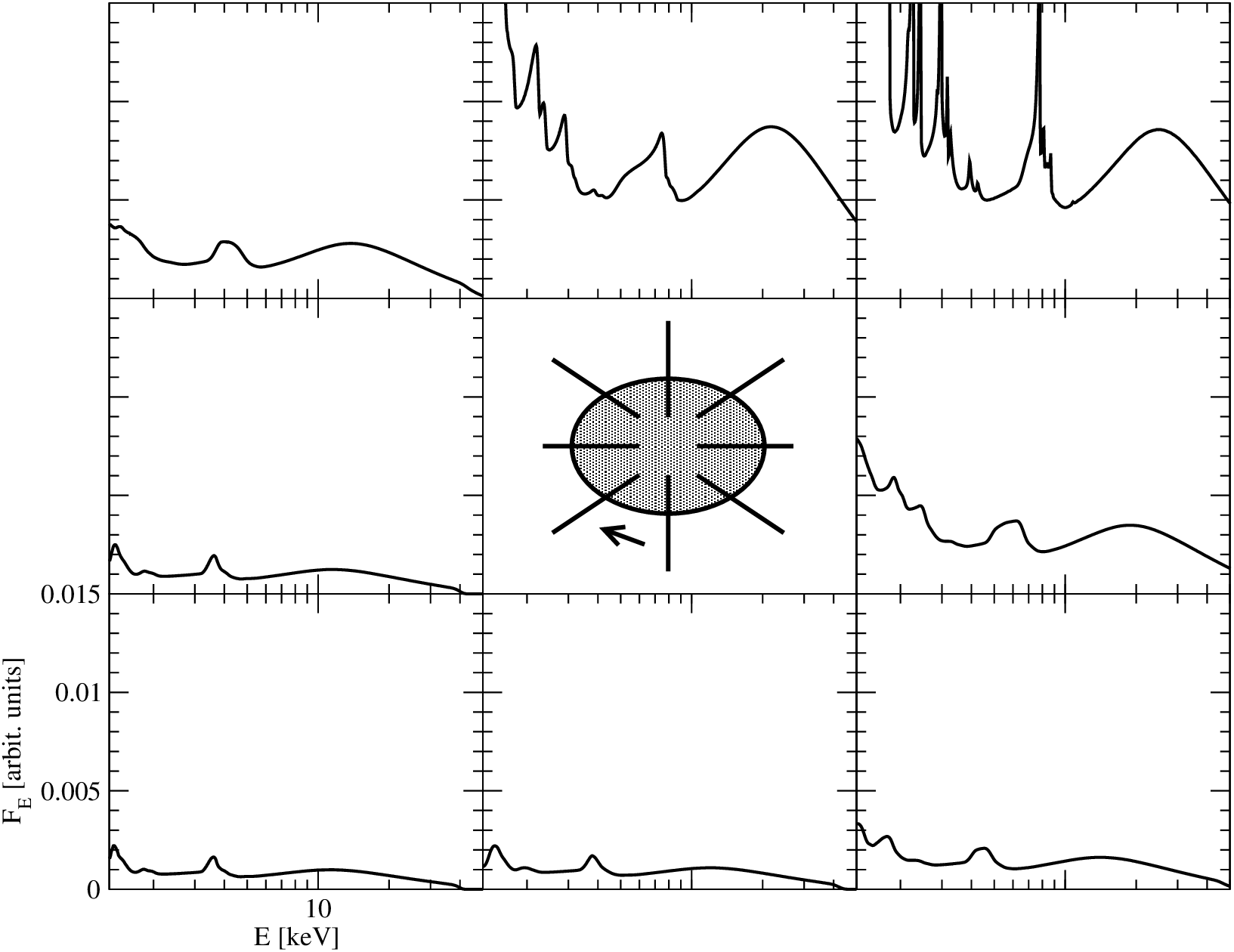}
  \hfill~
  \includegraphics[width=8.4cm]{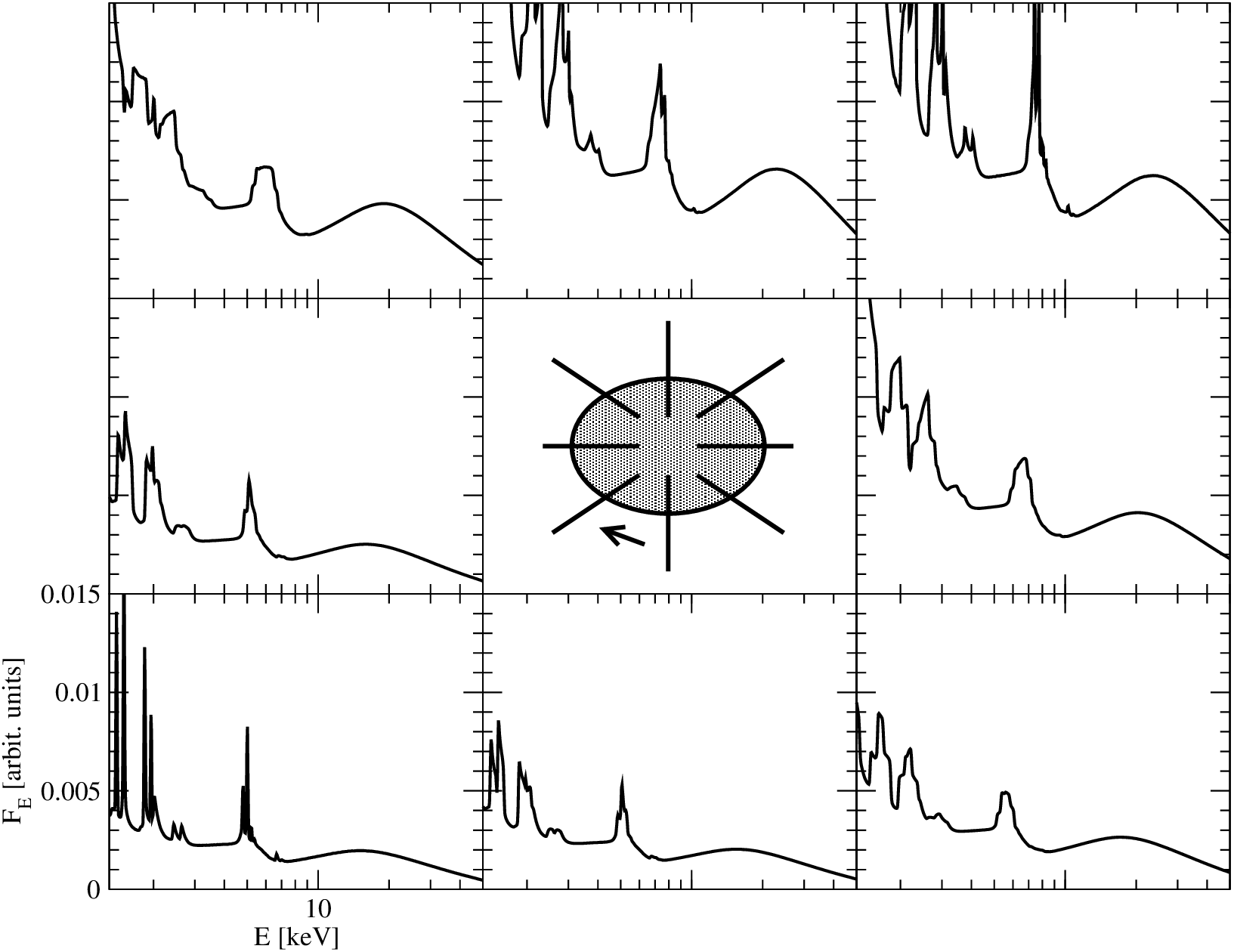}
  \vskip 0.3 truecm
  \includegraphics[width=8.4cm]{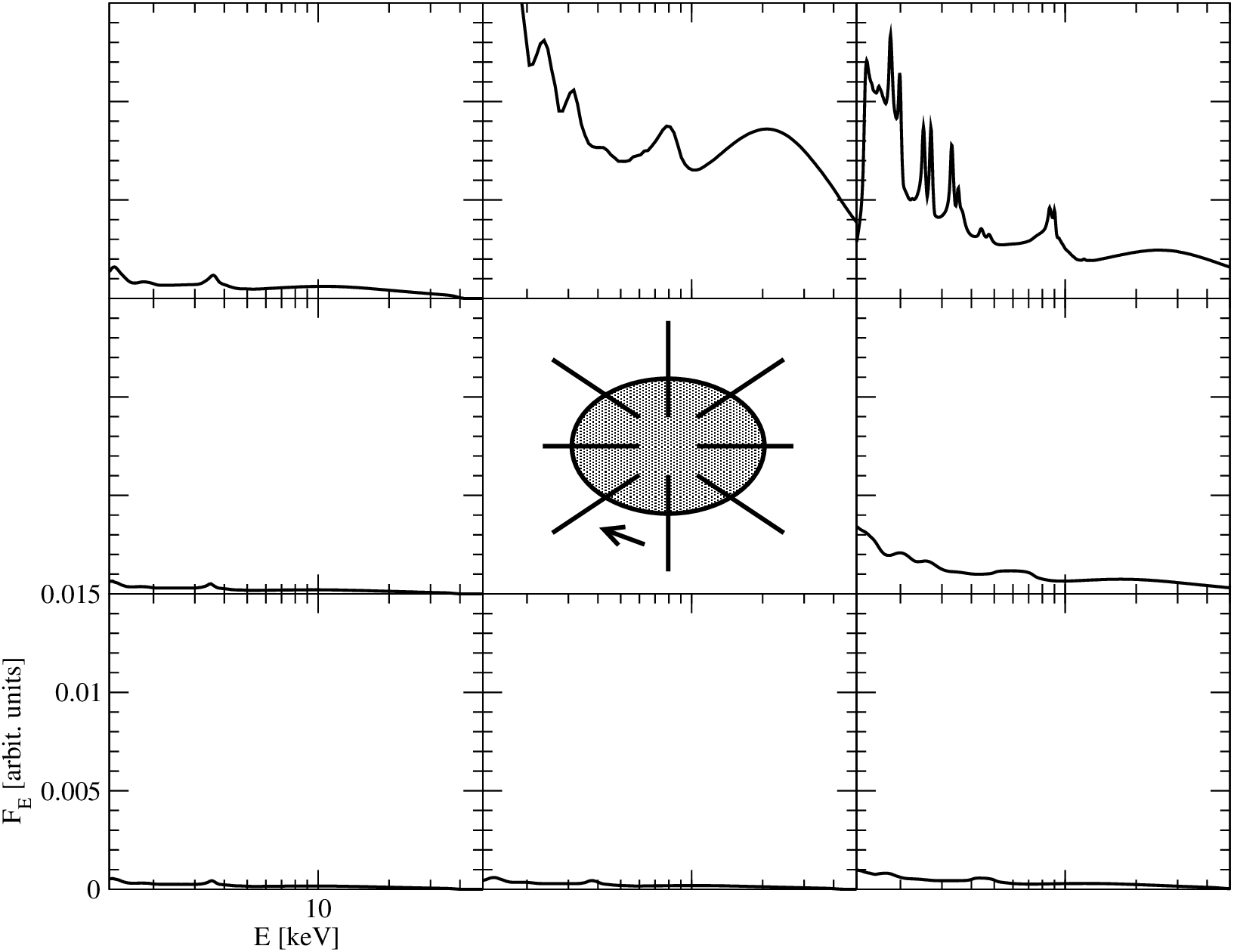}
  \hfill~
  \includegraphics[width=8.4cm]{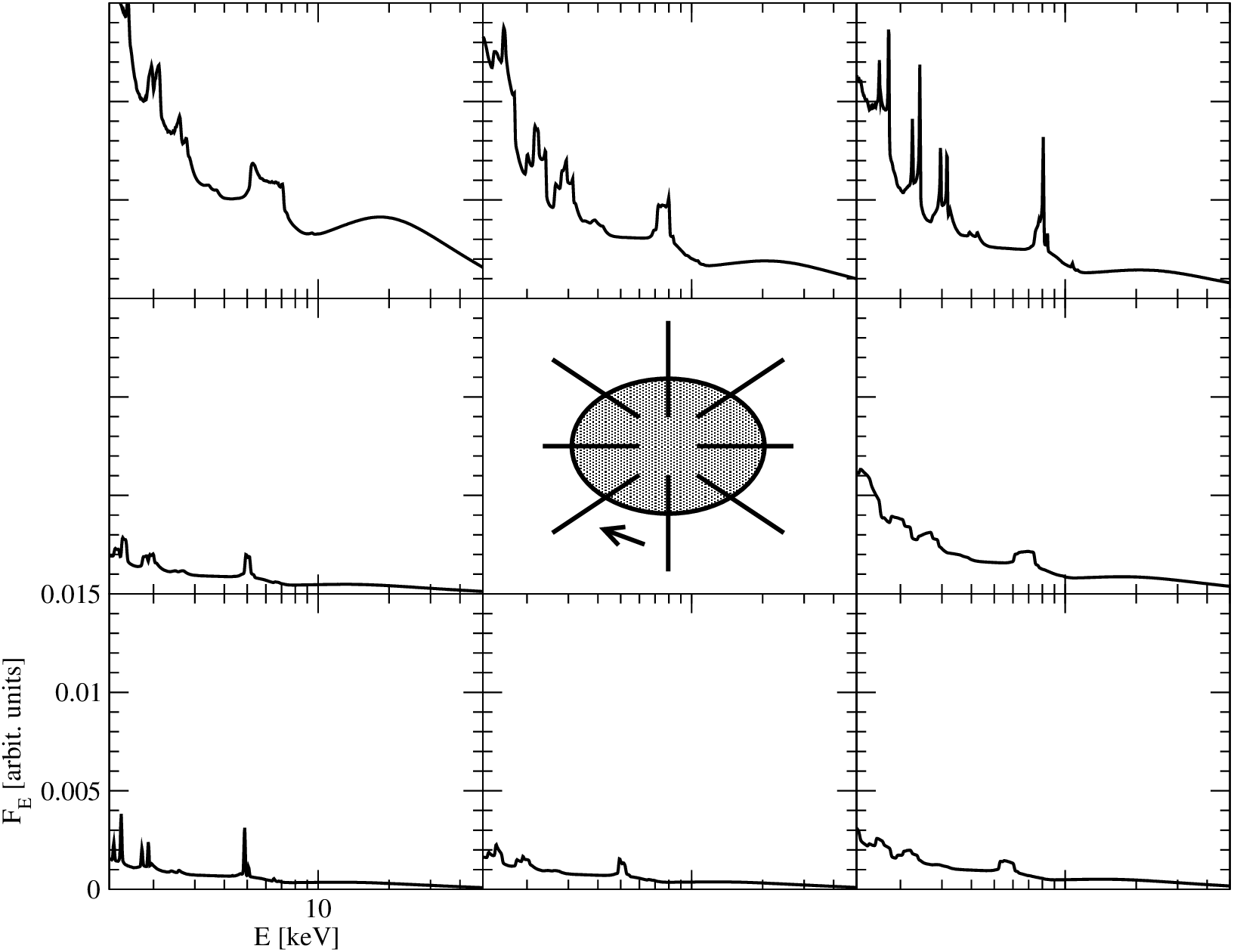}
  \caption{Spectra emitted by a persistent hot spot at 8 different orbital
    phases. In each panel the bottom middle spectrum corresponds to the starting
    phase of the spot's closest approach to the observer ($\phi_0 = 0\deg$). The
    spot exists for 1/8 of the orbital time scale at its location and the disk
    rotates clock-wise. From top to bottom the panels represent different viewing
    angles, $i$ (measured from the disk normal): $i = 30\deg$, $i = 60\deg$, and
    $i = 85\deg$. The spot is located at a distance of 7~$R_{\rm g}$ (left) and
    18~$R_{\rm g}$ (right) from the black hole. \label{fig:SM-eight-phase-spect}}
\end{figure*}

In Fig.~\ref{fig:SM-eight-phase-spect} we plot the spot spectra of a
persistent flare occurring at different orbital phases $\phi_0$. The
time-integrated spectra over (1--50)~keV are shown for three disk inclinations
with respect to a distant observer: $i = 30\deg$, $i = 60\deg$, and $i =
85\deg$ (top, middle and bottom panels respectively). The impact of general
relativity effects and of the spot motion becomes more important with
increasing $i$. It affects the normalization of the reprocessed spectra as
well as the shape of specific features. At both distances from the black hole
considered here, the relativistic modifications are relevant, but they are
more significant at 7~$R_{\rm g}$ than at 18~$R_{\rm g}$, as expected. The
spacetime curvature and the Doppler shifts are stronger when being closer to
the black hole. At 7~$R_{\rm g}$, the iron K-line complex cannot be properly
decomposed into components for any $\phi_0$ and $i$. An exception occurs at $i =
60\deg$, where shortly ``after'' the passage ``behind'' the black hole some
structure of the line appears. Still, the deformations remain quite strong
also in this case.

The spectral variations of a time-steady spot orbiting around a black hole are
not only due to relativistic Doppler shifting and gravitational lensing. To an
important extent they are also caused by time delay effects that are due to
variations in the ``length'' of different photon geodesics reaching the
observer at a given time in his/her reference frame
\citep{dovciak2007}. Gravitational lensing is especially important at edge-on
viewing directions, when the spot, the black hole, and the observer are
(nearly) lined up. At $i = 85\deg$, the maximum of the spectrum therefore
occurs when the spot passes ``behind'' the black hole, at $\phi_0 = 180\deg$
(top middle box in each panel). At intermediate inclinations the Doppler
shifting is more dominant. In contrast to the Newtonian case, the maximum
(minimum) {\it   relativistic} Doppler shift does not occur when the spot is
approaching (receding from) the observer at maximum velocity; it rather
appears around $\phi \sim 225\deg$ ($\phi \sim 45\deg$), which agrees with the
results shown in Fig.~\ref{fig:SM-eight-phase-spect}.

\subsubsection{Impact of the spot structure and of the angle-dependent local emission}
\label{sec:impact-effects}

We investigate the impact of the horizontal spot structure and the angle
dependent emission on the results. Therefore we compute time-integrated,
relativistically convolved spectra for the following three cases of the local
spot emission:

\begin{itemize}
  \item[\textbullet] including the detailed horizontal spot-structure and the
             $\psi$-dependence of the local spot emission,
  \item[\textbullet] averaging over the horizontal spot structure but keeping the
             dependence on $\psi$,
  \item[\textbullet] averaging over the horizontal spot structure and over $\psi$. 
\end{itemize}

We use the same model setup as for Fig.~\ref{fig:SM-eight-phase-spect}
and we define the specific case of $\phi_0 = 225 \deg$, i.e. the spot is on
the approaching side of the disk at the orbital phase where the strongest
relativistic modifications are expected. The spot then proceeds up to $\phi =
270\deg$. The inclination of the observer is set to $i = 60 \deg$. This may
correspond to a maximum type-1 viewing angle for very luminous AGN (quasars).

We obtain the following results: the differences between a horizontally
structured spot and a uniform one are negligible. The spot structure leaves a
trace in the spectral evolution only during the rising and fading
phases of the flare. These phases are very short compared to the total flare
life-time. The time-integrated spectrum is therefore dominated by the
reprocessed radiation coming from the fully illuminated spot, which evidently
can be well approximated by uniform emission. Accurate computations of the
angular emissivity dependence are slightly more important since they
significantly affect the normalization of the reflected component. Nevertheless, the
spectral shape is not strongly influenced: when the spectra are renormalized,
the differences between the cases a) and c) are less than 7\% at all energies.

We conclude that for a spot seen at $i = 60 \deg$ and orbiting at $R =
7$~$R_{\rm g}$ it is safe to simplify the computations for persistent flares by
averaging over the spot expansion. The same holds for lower inclinations, for
other orbital phases than $\phi_0 = 225 \deg$, and for higher values of $R$
because in all these cases the impact of the relativistic and Doppler effects
is smaller than for our test case.

\subsubsection{Detecting the orbital phase of a revolving spot}
\label{sec:orbit-phase}

In recent years, observations with {\it XMM-Newton} have started to reveal the
radial irradiation structure of the innermost regions in AGN accretion
disks. For the Seyfert galaxy NGC~3516 \citet{iwasawa2004} reported
time-dependent spectral data that are in agreement with the modeling of an
orbiting hot spot between 7~$R_{\rm g}$ and 16~$R_{\rm g}$. The data suggest a
periodic signal in the iron line flux, which can be interpreted as a long-term
flare completing several orbits. Nevertheless, the sensitivity of {\it
XMM-Newton} is not sufficient to clearly resolve sub-orbital features in the
spectral evolution of NGC~3516. Such investigations, which allow to map out
the azimuthal irradiation structure of the inner accretion disk, will become
possible with the much higher effective collecting areas of future satellite
observatories like {\it XEUS} and {\it Constellation-X}.

To illustrate the spectral appearance of sub-orbital X-ray signals we simulate
the observation of an orbiting hot spot applying our persistent flare
model. We adopt the response matrices of the {\it XEUS} {\sc TES} detector as
it is available in April~2007 and of the EPIC PN camera aboard {\it
XMM-Newton}. The simulations for both satellites are conducted in {\sc
xspec~v.11.3.2} using a modified version of the {\sc kylcr} model
\citep{dovciak2004a}. The model allows to implement our local spot spectra
described in Sect.~\ref{sec:local-prop} after they are averaged over the
horizontal spot expansion. According to Sect.~\ref{sec:impact-effects}, this
simplification does not induce spurious effects. With {\sc kylcr} we compute
the relativistically blurred, time-integrated spectrum of the spot that we use
to produce the simulated {\it XEUS} data. We define a persisting spot lasting
for 1/8 of an orbit at $R = 7$~$R_{\rm g}$ on the receding ($\phi_0 = 45
\deg$) and on the approaching ($\phi_0 = 225 \deg$) side of the disk. We
assume a Schwarzschild metric and a disk inclination of $i = 30\deg$. The
simulated photon flux of $5 \times 10^{-5}$~photons/(s~cm$^2$~keV) in the
variable red wing of the iron line complex is adjusted to the observation of
NGC~3516 analyzed in \citet{iwasawa2004}. We assume an observation time $T =
7300$~s, which corresponds to the flare life time assuming a black hole of
$10^8$ solar masses. The fake data are rebinned using the tool {\sc grppha} to
ensure that each bin contains a minimum of 50 (200) counts for the {\it
XMM-Newton} ({\it XEUS}) ``simulation''.

\begin{figure}
  \centering
  \vskip 0.1cm
  \includegraphics[width=\columnwidth]{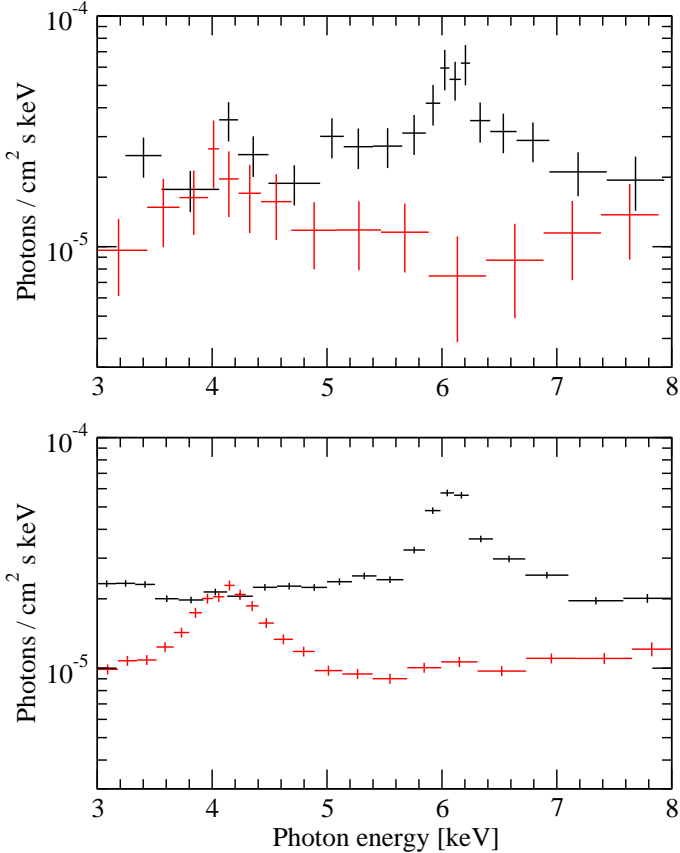}
  \caption{Simulated observations by {\it XMM-Newton} (top) and {\it XEUS}
    (bottom) of a relativistically broadened iron K-line emitted by localized
    X-ray sources that orbit at $7 \, R_{\rm g}$ around a $10^8$ solar mass
    black hole. The two hot spots are at opposite positions, on the receding
    ($\phi_0 = 45\deg$, red) and the approaching ($\phi_0 = 225\deg$, black)
    side of the accretion disk.  
    \label{fig:XEUS-fake}} 
\end{figure}

In Fig.~\ref{fig:XEUS-fake} we plot the produced spectra. For {\it XEUS} the
flares on the two sides of the accretion disk can be distinguished from
each other as the normalization and the position of the iron K-line differ
significantly. Also in the {\it XMM-Newton} spectrum these differences are
visible but the much higher noise level does not allow the same clear
conclusion. Note that in these simulations we assume that the primary
radiation does not dilute the reflection. This holds true only if the object
is strongly reflection-dominated or if the primary radiation has been precisely
removed in the data analysis. In the latter case additional errors are
expected, which makes it even less likely that {\it XMM-Newton} resolves such
sub-orbital features.

\subsubsection{Probing the local spot spectrum}

The computations of the local spectra we use include the vertical disk
structure and the radiative transfer in such a stratified medium. The
resulting reprocessed spectra reveal several iron line components, which are
then blurred by the relativistic effects. The subtleties  of the resulting
spectrum cannot be well resolved in present-day X-ray data, and the question
arises whether they will be visible with more advanced instruments in the
future. We investigate to which extend the spectral structure in the iron line
band can be observationally resolved by comparing simulated data from {\it
XMM-Newton} and {\it XEUS} for a persistent flare. 

\begin{figure}
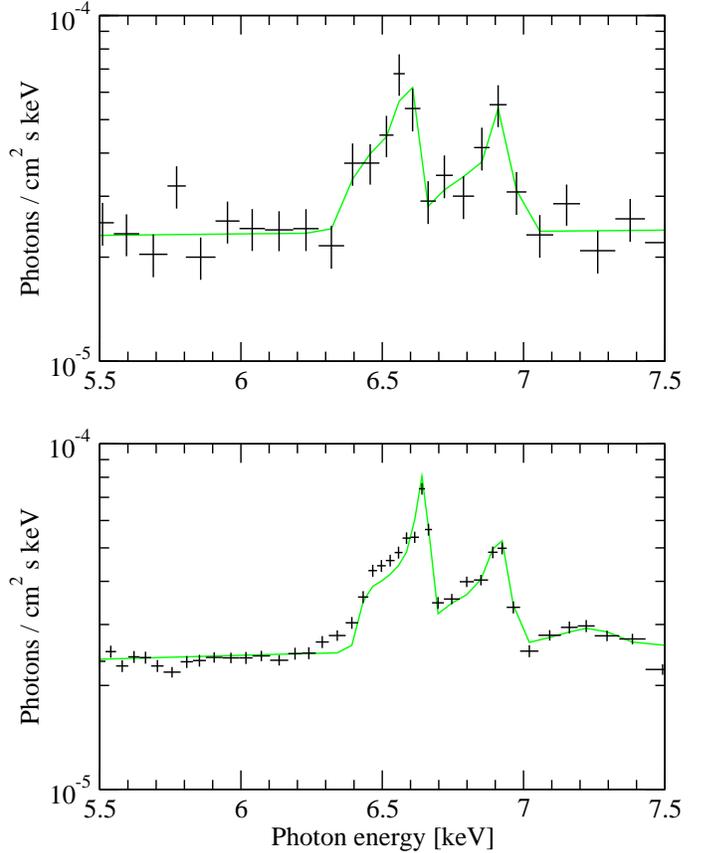

  \vskip 0.1cm
  \centering
  \includegraphics[width=\columnwidth]{8273fg13.eps}
  \includegraphics[width=\columnwidth]{8273fg14.eps}
  \caption{Best fit for the simulated {\it XMM-Newton} (top) and {\it XEUS}
    (bottom) data of an approaching hot spot orbiting at $18 \, R_{\rm g}$
    around a $10^8$ solar mass black hole. The best fit models are indicated
    by solid lines: ``power law + 2 {\sc kygline} components'' for the {\it
    XMM-Newton} data and ``power law + 3 {\sc kygline} components'' for the {\it
    XEUS} simulations. \label{fig:1fake}}
\end{figure}

We consider a hot spot lasting for 1/8 of the orbital period at $R = 18 \,
R_{\rm g}$ corresponding to an integration time of 30700~s for a $10^8$~solar
mass black hole. At this disk radius from the black hole, the relativistic
effects are still relevant but not as violent as close to the marginally
stable orbit. We therefore expect to still detect some structure of the iron
line in the fake observations. The flare is supposed to start on the
approaching side of the accretion disk, at $\phi_0 = 225\deg$, and  the system
is seen at an inclination of $i = 30\deg$. The simulated observations are then
obtained as described in Sect.~\ref{sec:orbit-phase}. They are shown in
Fig.~\ref{fig:1fake}.

For the purpose of data analysis, it is a common contemporary approach to
relativistically blur one or several narrow Gaussians to approximate the local
profile of the iron line, whereas the continuum is modeled by an underlying
power law (which is invariant under the relativistic effects). We investigate
how well the simulated data of a persistent spot can be reproduced by such
phenomenological models. We fit the simulated data consecutively adding more
relativistically broadened line components to the underlying continuum power
law. The lines are implemented by the {\sc kygline} that is also available
in {\sc xspec}. The disk emission pattern is defined exactly as for the data
simulations, but the parameters of the power law as well as the line centroid,
width, and normalization are allowed to vary in the fitting process. The
fitting is restricted to the spectral range of (4.0--7.5)~keV to avoid the
soft X-ray features and the strong iron absorption edge.

As expected the resulting phenomenological fits are much better determined for
the more precise {\it XEUS} data than for the {\it XMM-Newton} simulations
(Fig.~\ref{fig:1fake}). For {\it XEUS} it is possible to find a
phenomenological fit for three distinguished components of the K$\alpha$-line
complex (Table~\ref{tab:fake}): neutral/weakly ionized iron ($\sim 6.4$~keV),
helium-like iron ($\sim 6.7$~keV), and hydrogen-like iron ($\sim
6.9$~keV). The attempt to also identify an individual K$\beta$-line around
7.1~keV fails, instead the hydrogen-like component seems particularly broad
indicating a possible blend. All line centroids are slightly redshifted with
respect to the laboratory centroid energies, also the hydrogen-like
K$\alpha$/neutral K$\beta$ line when interpreted as a blend. Visually the
residuals imply that the fit misses some curvature in the spectrum red- and
bluewards of the broadened iron line. The intrinsic spectral shape across the
iron-line band is more complex than a power law continuum plus
relativistically blurred Gaussian lines can account for. In the local
reprocessing spectra, the various components of the iron K-line complex seem
to stand on a ``pedestal'' between 6~keV and 7~keV (see
Fig.~\ref{fig:r3-i4-Ballan-spect}). This broad feature is mainly due to the
drop in continuum normalization at the iron K absorption edges above
7.1~keV. The edges should therefore be considered in the fitting. Redward of
the 6.4~keV line, some impact of a Comptonized line shoulder might also be
relevant.

\begin{table}
  \centering
  {\scriptsize
  \begin{tabular}{llccc}
    \hline
    \noalign{\smallskip}
    Model component & Parameter & {\it XEUS} & {\it XMM-Newton} & {\it XMM-Newton}\\ 
    \noalign{\smallskip}
    \hline
    \noalign{\smallskip}
       Power law    & $\Gamma$                & -0.303                & -0.112               & -0.091 \\
                    & $N$                     & 1.42$\times 10^{-05}$ & 1.91$\times 10^{-05}$ & 1.96$\times 10^{-05}$\\
    \noalign{\smallskip}
       {\sc kygline} \#1 & $E_{\rm c} [\rm{keV}]$  &  6.381               & 6.340                 & 6.458 \\
                   & $\sigma [\rm{eV}]$      & 12.713               & 20.243                & 180.95 \\
                   & $N$                     & 6.06$\times 10^{-06}$ & 6.44$\times 10^{-06}$ & 1.09$\times 10^{-05}$\\
    \noalign{\smallskip}
       {\sc kygline} \#2 & $E_{\rm c} [\rm{keV}]$  &  6.647                &  6.659               &     --     \\
                   & $\sigma [\rm{eV}]$      & 21.931                & 0.115                &     --     \\
                   & $N$                     & 4.19$\times 10^{-06}$ & 4.35$\times 10^{-06}$ &     --     \\
    \noalign{\smallskip}
       {\sc kygline} \#3 & $E_{\rm c} [\rm{keV}]$  &  6.983                &     --     &     --     \\
                   & $\sigma [\rm{eV}]$      & 59.741                &     --     &     --     \\
                   & $N$                     & 8.85$\times 10^{-07}$ &     --     &     --     \\
    \noalign{\smallskip}
    \hline
    \noalign{\smallskip}
                    & Red. $\chi^2$        & 352/250  &  27/33  & 39/36 \\
    \noalign{\smallskip}
    \hline
  \end{tabular}
  }
  \vskip 0.2cm
  \caption{Best-fit parameters for simulated {\it XEUS} and {\it XMM-Newton}
    observations of a hot spot at the phase of approaching motion. Power law
    models combined with up to three  relativistically blurred iron lines are
    compared.}
  \label{tab:fake}
\end{table}

In the case of {\it XMM-Newton}, the number of required sub-components is
much less well-defined (Fig.~\ref{fig:1fake}). Models with one or two Gaussian
components both lead to acceptable fits (Table~\ref{tab:fake}), but including
both, a neutral/weakly ionized and a helium-like K$\alpha$-component deliver
the slightly better fit. For the model that only consists of one line the 
line width is very large as expected. In the two-line model only the
neutral/weakly ionized iron line is broad and the helium-like component
appears very narrow. Any further line around (6.9--7.1)~keV remains undetected.

An important result of the modeling is, however, that repeating the same data
simulations for {\it XMM-Newton} shows a significant uncertainty in the
fitting results and their interpretations. The results for {\it XEUS} on the
other hand are stable. This documents that the sensitivity of {\it XMM-Newton} is
not high enough to draw reliable conclusions about the sub-structure of the
broad iron line shape coming from a persistent flare at $18 \, R_{\rm g}$ as it
is considered here.

\subsection{Transient spots at different orbital phases}
\label{sec:flare-evo}

In a second step we investigate the reprocessed spectra of flares with a
life-time that is comparable to the light-crossing time of the spot. We name
such spots as {\it transient} and we investigate their emission at defined
orbital phases. The irradiating primary emission is switched on and off
instantaneously. The emission lasts for one gravitational time scale, $T_{\rm
g} = \frac{GM}{c^3}$, which compares to the spot's light-crossing time by
$T_{\rm g} \sim 1.15 \, \frac{r_0}{c}$. During $T_{\rm g}$ the spot proceeds
by the azimuthal angle $\Delta \phi (R) = (R_{\rm g}/R)^{1.5}$. At the two
disk radii considered here this amounts to $\Delta \phi (7 \, R_{\rm g}) \sim
3\deg$ and $\Delta \phi (18 \, R_{\rm g}) \sim 0.75\deg$. Hence, during the
flare irradiation the spot remains nearly at the same orbital phase. For the
initial azimuthal position $\phi_0$ of the flare we consider the values
$\phi_0 = 0\deg$, $\phi_0 = 90\deg$, $\phi_0 = 180\deg$, and $\phi_0 =
270\deg$. We use the same radiative transfer solutions for the local
reprocessed spectra as in Sect.~\ref{sec:local-prop}.  We then run {\sc KY} to
compute sets of relativistically blurred, time-dependent spectra at the
viewing angles $i = 30\deg$, $i = 60\deg$, and $i = 85\deg$.

\subsubsection{Spectral appearance}

\begin{figure*}
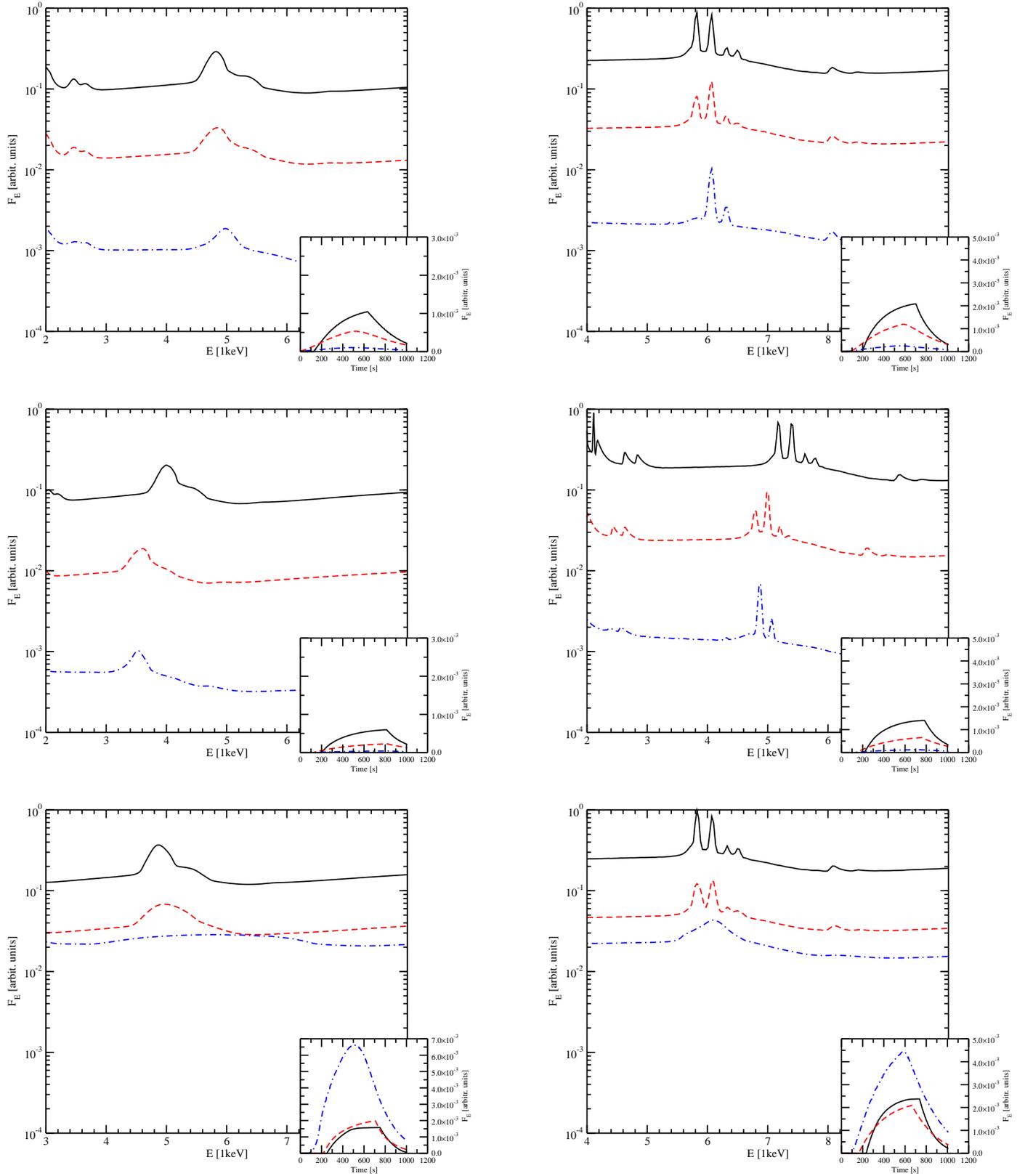

  \centering
  \includegraphics[width=8.5cm]{8273fg15.eps}
  \hfill
  \includegraphics[width=8.5cm]{8273fg16.eps}
  \vskip 0.5 truecm           
  \includegraphics[width=8.5cm]{8273fg17.eps}
  \hfill                      
  \includegraphics[width=8.5cm]{8273fg18.eps}
  \vskip 0.5 truecm
  \includegraphics[width=8.5cm]{8273fg19.eps}
  \hfill                     
  \includegraphics[width=8.5cm]{8273fg20.eps}
  \vskip 0.1 truecm
  \caption{Integrated flux spectra and lightcurves for short-term flares at
    four different initial azimuthal positions of the spot. From top to
    bottom, the panels represent $\phi_0 = 0\deg$, $\phi_0 = 90\deg$, $\phi_0 =
    180\deg$, and $\phi_0 = 270\deg$ (continued on the next page). The origin
    of the azimuthal scale at $\phi = 0\deg$ is at the closest approach of the
    hot spot to the (inclined) observer; for $\phi = 180\deg$ the spot is
    farthest away and ``behind'' the black hole.The spectra were integrated
    over the entire flare period. The various curves of one panel denote the
    disk inclinations $i = 30\deg$ (black, solid), $i = 60\deg$, (red,
    dashed), and $i = 85\deg$ (blue, dot-dashed). The spectra contain a
    systematic offset by a factor of 5 between each other for clarity. Results
    for a spot at $R = 7 \, R_{\rm g}$ (left panels) and at $R = 18 \, R_{\rm
    g}$ (right panels) from the central black hole are
    shown. \label{fig:flash-7rg}} 
\end{figure*}

In Fig.~\ref{fig:flash-7rg} we present the obtained integrated spectra for the
observed flare periods and the corresponding lightcurves at the different
azimuthal positions. The observed flare durations are of the order of a few
hundreds of seconds. The plots show that line broadening is important also for
transient flares. But in contrast to the persistent spots the spot velocity
projected along the observer's line-of-sight does not change significantly
during the flare period; this holds true even at high disk inclinations as the
spot only moves by very small values of $\Delta \phi$. We therefore conclude
that the line broadening is caused to a large extent by general
relativity. Due to the expansion of the spot, the photons arriving at the
distant observer at a given time originate at different positions on the
accretion disk. Hence, their gravitational redshift, which only depends on the
distance of the emission point from the black hole, is distributed over a
certain range and the line appears to be broadened. At 18~$R_{\rm g}$ (right
panels of Fig.~\ref{fig:flash-7rg}) the gravitational line broadening is
weaker and the iron K-line complex is resolved into its sub-components at all
orbital phases. When comparing the iron K-lines for the same $\phi_0$ but at
different inclinations, the effects of the vertical ionization and temperature
profile of the disk become visible (see Sect.~\ref{sec:local-prop}) and
indicate that toward higher inclinations the medium appears to be more
strongly ionized. 

\begin{figure*}
  \centering
  \includegraphics[width=8.5cm]{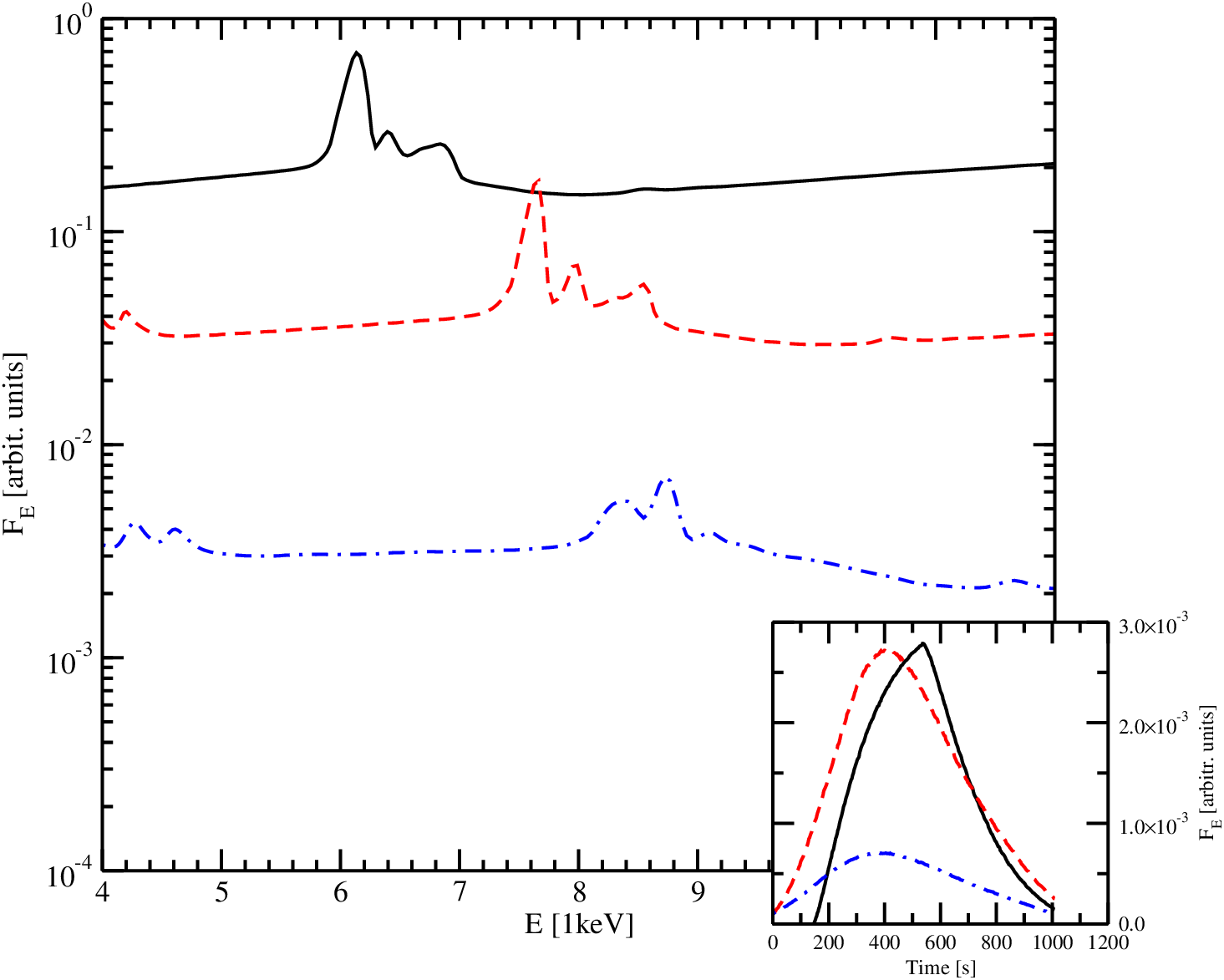}
  \hfill
  \includegraphics[width=8.5cm]{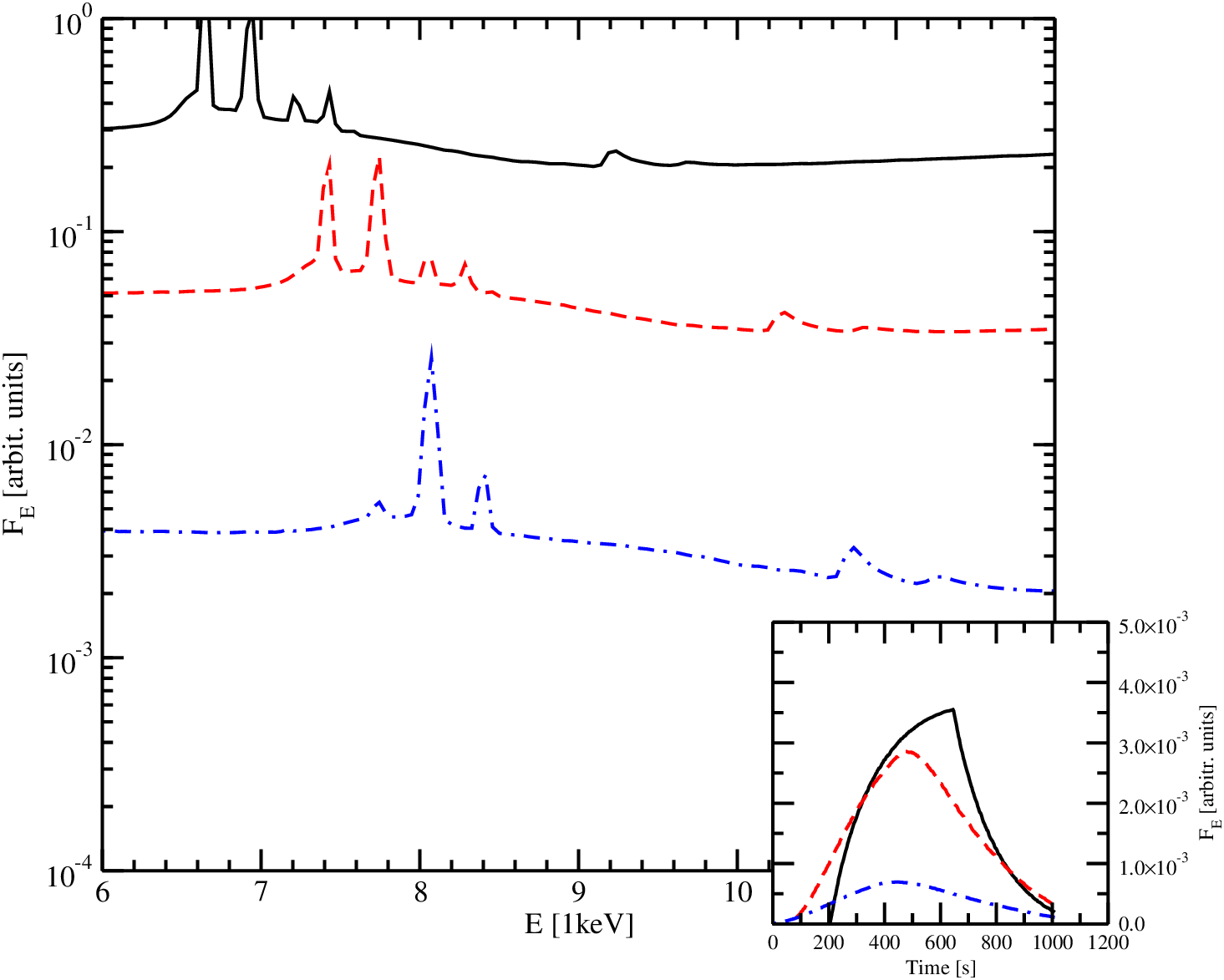}
  \vskip 0.1 truecm
  Fig.~7. -- continued
\end{figure*}

Some aspects of the spectral shape are consistent with the results obtained in
Sect.~\ref{sec:SM-whole-orbit} for a persistent hot spot completing a
significant fraction of the orbit. The Doppler shift of the K$\alpha$ line
complex, for instance, corresponds to the orbital phase of the flare. It
increases with higher inclination due to a larger velocity component projected
along the line-of-sight. For a flare occurring ``behind'' the black hole,
i.e. for $\phi_0 = 180\deg$, the lensing effect strongly modifies the spectrum
and increases the overall flux. At this phase the iron K-line is entirely
smeared out when seen at high inclinations.

\subsubsection{Lightcurves}

In our modeling of transient flares we consider the different light-travel
times between the elevated source and various positions inside the hot
spot. The effect can be seen in the lightcurves shown in
Fig.~\ref{fig:flash-7rg}. The lightcurves represent the integrated flux
between 2~keV and 50~keV. The \emph{observed} flare period is not the same for
different values of $i$ and $\phi_0$. Physically, it depends on the difference
between the ``longest'' and the ``shortest'' photon geodesics connecting the
spot with the distant observer at a given time in the observational frame. The
geodesics are a function of the spacetime geometry and the photon flux they
transfer depends on the time-dependent emission across the spot.

At low inclinations the lightcurves have a characteristic and regular shape
showing the rising and fading of the flare emission. The shape is not at all
box-like as assumed for the incident radiation illuminating the spot. The
geometrical expansion of the spot induces different time-delays for the
reprocessed radiation and so the onset and the fading of the spot emission are
smoothed out. A significant time asymmetry is seen in most cases, with a
shallower rise and a more rapid decay. Going toward higher inclinations, the
shape of the light curve changes and generally becomes more symmetric -- with
an exception for the phase $\phi_0 = 180\deg$, when the spot passes ``behind''
the black hole. The closest symmetry is reached at $\sim 60\deg$. For even
larger $i$ the flux decreases rapidly and the lightcurves become much flatter.

\subsubsection{Spectral time evolution}

\begin{figure*}
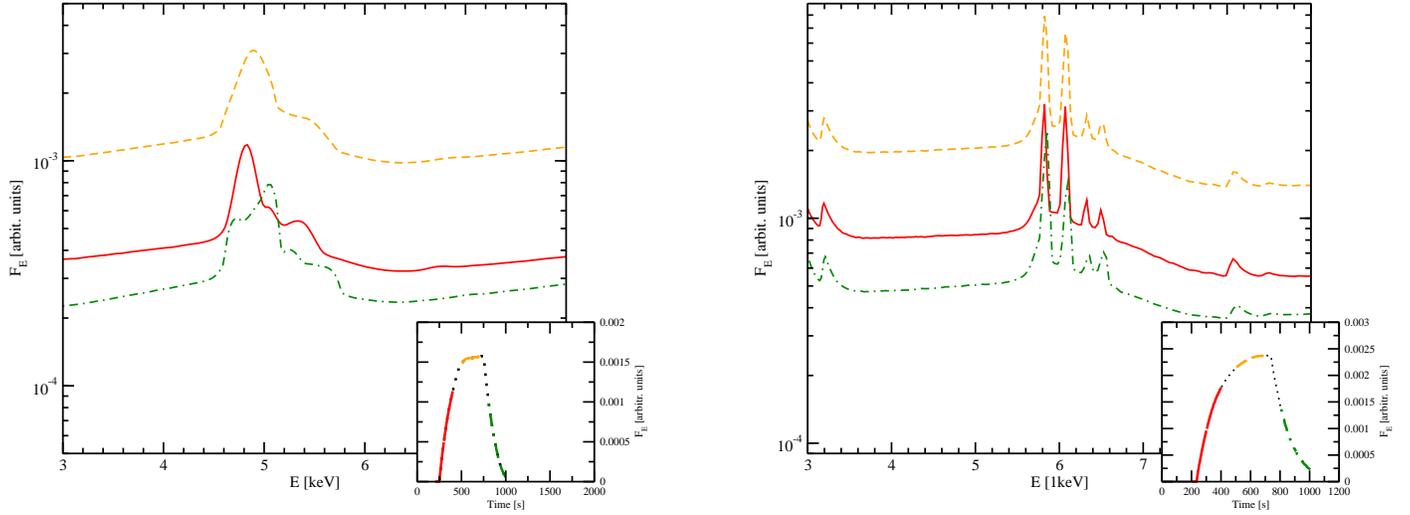

  \centering
  \includegraphics[width=8.5cm]{8273fg23.eps}
  \hfill                      
  \includegraphics[width=8.5cm]{8273fg24.eps}
   \caption{Spectral evolution and lightcurves for a transient flare located
     ``behind'' the black hole ($\phi_0 = 180\deg$) with respect to the
     observer. The observer's inclination is $i = 30\deg$. The flare periods are
     divided into separate time intervals, over which the spectra are
     integrated. The corresponding time interval of each spectrum is marked on
     the flare lightcurve. Results for a spot at $R = 7 \, R_{\rm g}$ (left panels)
     and $R = 18 \, R_{\rm g}$ (right panels) are shown.\label{fig:flash-spec-evo}}
\end{figure*}

To investigate the spectral evolution during a short flare we present in
Fig.~\ref{fig:flash-spec-evo} snapshots of a time-dependent spot spectrum. For
the particular model shown we consider a spot passing ``behind'' the black
hole ($\phi_0 = 180\deg$) and a disk inclination of $30\deg$, which is typical
for type-1 AGN.

The development of the flare emission across the hot spot is visible in the
spectral sequence. The irradiation sets on at the spot center where it
produces the highest temperature and ionization states. Therefore, the iron
K-line complex shows stronger components of helium- and hydrogen-like iron for
the rising branch of the lightcurve. During the flare period these components
become weaker and the K$\alpha$ and K$\beta$ emission of less ionized/neutral
iron increase. The development of the line components is more clearly visible
for a spot at $R = 18$~$R_{\rm g}$. At 7~$R_{\rm g}$, i.e. close to the
marginally stable orbit, the relativistic broadening of the line makes it
difficult to distinguish between its individual components. Nevertheless, a
development of the line from the onset to the decay of the flare can be
seen. The spot evolves from a small emission site at its center to a
circle/ring with outer radius $r_0 = 0.866$~$R_{\rm g}$. Hence, the gradient
in gravitational and Doppler shifting across the illuminated spot increases
during the spot's life-time, so that the width and the centroid of the iron
K-line evolve.

\section{Importance of the studied effects}
\label{sec:effects}

The spot model we present in this paper contains accurately computed local
spectra for the reprocessed flare emission. We include the vertical
stratification of the accretion disk and the geometrical details of
the flare reprocessing before applying the relativistic modifications and
computing the observed spectra. In previous work, these effects have not all been
considered. We discuss in this section their impact on the results and we
estimate to which extend it is necessary to include them into the modeling.

\subsection{Local emission from the spot}
\label{sec:loc-emi}

We calculated the local spot spectrum taking into account the vertical
density stratification of the disk. Our local spectra exhibit more than
one strong component of the iron line and a complex reprocessed continuum. The
continuum component includes a series of iron K absorption edges above
7.1~keV and a Comptonized shoulder on the red side of the K$\alpha$-line at
6.4~keV. These features give rise to a broad continuum ``pedestal'' carrying the
emission lines, which is not completely smeared out by the relativistic
effects even close to the marginally stable circular orbit of a
Schwarzschild black hole.

From simulating future {\it XEUS} observations of persistent flares in AGN at
$18 \, R_{\rm g}$, we conclude that the sub-structure of the iron line complex
can be investigated probing the reflecting medium. Using a detailed reflection
model for the data analysis instead of phenomenological models has the
advantage that the particular spectral characteristics are consistently
accounted for. The normalization of the emission lines with respect to the
continuum stays on physical grounds and can be connected to global parameters
of the system (e.g. ionization parameter, disk radius). For the case of {\it
Suzaku} observations of AGN the Compton hump around 30~keV can be fitted
coherently with the broad iron K-line complex as both features are due to
reprocessing \citep[e.g.][]{george1991}.

Careful computations of the radial structure of a spot are not necessary
if the time integration of the data is longer than the light travel time
across the spot. In this case the spot spectrum can be approximated as a
uniform (average) spectrum  which simplifies the calculations. Since the
flares considered in the current paper are located at a small height
above the disk, the spot radii are small and this assumption is likely to be
satisfied in most cases. The radial spot structure is going to be  more
important for flares located at large heights above the disk, but in this
case the flare is a non-local phenomenon, and both the disk radial
structure as well as the light bending between the flare and the disk
surface might be important. Such a model becomes essentially identical
to the lamp-post geometry, which is discussed in Sect.~\ref{sec:LP}.

\subsection{General relativity and Doppler effects}

We have included photon energy shifts due to Doppler and general relativity
effects, as well as light bending and mutual time delays that influence the
observed signal. These effects have a strong impact on the observed
spectral features and should be taken into account when modeling reprocessed
radiation at distances smaller than 20~$R_{\rm g}$ from the black hole. We
find, that the relativistic effects do not only affect the observed shape of
the iron K-line emission but also the underlying continuum ``pedestal'' that
appears in the iron line band (see Sect.~\ref{sec:loc-emi}).

The relativistic effects are important for the localization of the
hard X-ray emission. Short exposure time spectra, which cover only a
fraction of the Keplerian timescale of the co-rotating flare, clearly
exhibit the dependency of the observed spectral shape on the spot
position. The observer's viewing angle is also important, as it determines
the line-of-sight motion of the disk material, i.e. the Doppler shifts.
At very high inclinations the gravitational lensing dominates the
relativistic effects at the moment when the hot spot passes behind the black
hole. This can be a relevant case for Compton-thin type-2 AGN.

\subsection{Flare model versus lamp-post model} 
\label{sec:LP}

The flare model is a viable scenario for the primary X-ray emission in objects
with relativistically broadened iron lines. Multiple flares are expected if
the dynamo action leads to the occasional emergence of magnetic loops filled with
plasma above the disk surface. Another popular model assumes that a single
variable source is located near the symmetry axis; it could be tentatively
identified  with a shock at the jet base (Henri \& Pelletier 1991). The
existence of a single ``lamp post'' implies that the loop formation is
suppressed and a collimated outflow is considered instead. The two schemes differ
rather deeply in the underlying physics (Martocchia et al. 1996, 2000;
Miniutti \& Fabian 2004).

Both geometries currently have observational support. A single source at
variable heights is able to explain the apparent lack of variability of the
line expected to accompany the variations of the primary (Miniutti \& Fabian
2004; Larsson et al. 2007; but see \. Zycki \& Niedzwiecki 2006 for a critical
analysis of the model). The multiple flare scenario is supported by detections
of powerful irradiation occasionally illuminating the disk at large radii
\citep{turner2006}, and in the case of many co-existing flares
the scenario can also explain the apparent suppression (Czerny et al. 2004)
and the relative lack of variability \citep{goosmann2006a} of the iron K-line.

Obviously the two geometries share common properties, if a lamp-post source is
off-axis (Dabrowski \& Lasenby 2001), but differences should be detectable in
their time-dependence. The light traveling times between the flare source and
the reprocessing disk are significantly larger for a lamp-post geometry than
for the flare geometry assumed in this paper. Such delays can be constrained
from cross-correlation analysis between the lightcurves at different energy
bands as shown in \citet{goosmann2007}. Furthermore, we have shown in
Sect.~\ref{sec:orbit-phase}, how the more accurate spectroscopic observations
expected from future satellites such as {\it XEUS} and {\it Constellation-X}
will enable to map out the azimuthal details of the irradiated accretion
disk -- the azimuthal pattern of the disk re-emission depends on the radial
position of the flare source, and is symmetric only when the flare is located
on the disk axis, i.e. in the strict lamp post case.

\section{Discussion and conclusions}
\label{sec:discuss}

We present exemplary results from very advanced time-dependent modeling of the
reprocessed emission by a hot spot created underneath a magnetic flare. We
attempt to test the sensitivity of these models to various simplifying
assumptions. This gives the basis to determine a suitable grid of local
solutions, which is needed to model an arbitrary flare with reasonable
accuracy and efficiency, having in mind future X-ray missions like {\it XEUS}
and {\it Constellation-X}. Our spot model includes the time-dependent effects
that are caused by the delayed reprocessing from various parts of a spot. We
have also included the horizontal spot structure and the angular dependence of
the local emission. The local emission is modeled taking into account the
vertical disk structure and the observed time-dependent spot spectrum contains
all relativistic effects. We summarize the most relevant results we obtain:

\begin{itemize}

\item[\textbullet] The vertical temperature profile of the disk underneath the
  hot spot and the resulting local reprocessed spectra vary significantly
  across the spot surface. The ionized skin of the disk is thicker and the
  spectrum softer at the spot center than at the spot border. The different
  components of the iron K-line also follow this temperature gradient.\\

\item[\textbullet] At all positions inside the spot, the local spectra show a
  limb-brightening effect and spectral softening with increasing emission
  angles $\psi$. The iron line complex develops toward higher ionization
  states as $\psi$ rises.\\

\item[\textbullet] The relativistically modified spectra seen by a distant
  observer depend significantly on the flare position. As expected, the
  spectral distortion of the iron K-line becomes more important at smaller
  disk radii and at higher disk inclinations.\\

\item[\textbullet] For persistent flares the radial dependence of the local
  spectra across the hot spot can be safely averaged over, but the dependence
  of the local emission angle still has a small impact on the resulting
  spectrum seen by a distant observer.\\

\item[\textbullet] Current observational techniques are only at the threshold
  to determine the azimuthal phase of persistent hot spots. Their detection
  would require a perfect subtraction of an eventual primary component.\\

\item[\textbullet] Transient spots show a slight spectral development reflecting the
  evolution of the local emission from the spot center toward the border. They
  reveal a large variety of shapes in X-ray lightcurves depending on the spot position
  and the observer's inclination.

\end{itemize}

Especially the horizontal time-dependent re-emission of the transient spot has not
been assessed in previous papers. It turns out that it has a significant
impact on the observed lightcurve of the reprocessed radiation. The rising and
the decaying phase have characteristic shapes and they do not look symmetric at
disk inclinations smaller than $60\deg$. This result will be testable by
future observation techniques with a high-resolution and a high
time-throughput. On the other hand, we obtain only minor differences in the
spectral time-evolution between the rise and the decay of the spot
emission. Even with {\it XEUS} or {\it Constellation-X} they should remain
undetected as the maximum observation time is set by the flare duration. In
our case we assume it to be 1~$T_{\rm g}$, equating only a few hundreds of
seconds for a black hole with $10^8$~$M_\odot$.

The relevant time-scale of a transient flare as we assume it in this work
scales with the black hole mass. One could therefore imagine that the
spectroscopic analysis of transient flares should become possible for higher
black hole masses observed in some quasars. But unfortunately, high mass
quasars tend to be more distant and therefore their count rate in the X-ray
range is low compared to nearby, lower-mass Seyfert galaxies. Longer
observable delay times between the primary and the reprocessed radiation are
rather expected when the flare source is located higher above the disk than $H
= 0.5$~$R_{\rm g}$ as assumed in our model. But such a modeling approach
requires to also include the relativistic effects for the intrinsic radiation,
which could be neglected for the small value of $H$ we considered.

The situation is much more promising for persistent flares. Using {\it XEUS}
or {\it Constellation-X} to observe persistent spots in AGN the obtained data
will be accurate enough to constrain not only the azimuthal location
from X-ray spectroscopy (see Sect.~\ref{sec:orbit-phase}), but it can
even give some hints on the properties of the local spectrum. We find that the
horizontal spot structure has practically no impact on the observed spectrum
so that the spot can be approximated by a uniform average
emissivity. Nevertheless, this simplification would no longer be allowed for
spots being close to the marginally stable orbit of a rotating black hole. In
this case the relativistic modifications are more extreme and vary more
strongly across the spot surface.

In our current model, we have restricted ourselves to the case of flares
that occur near above the disk ($h\sim 0.5 \, GM/c^2$) and relatively
far from the black hole ($r\gtrsim 6 \, GM/c^2$), so that the light-bending of
the primary X-rays can be neglected. The primary reason for this restriction
was motivated by our goal to concentrate on the properties of the spot
spectra in short-exposure observations. This restriction means that we
do not consider the influence of the black hole rotation, which would
further complicate the analysis. Of course, a generalization to the rotating
(Kerr) case should be explored in the future, because there are good chances
that short-exposure spectra will remove the degeneracy which affects
time-integrated spectral profiles. (There are no principal difficulties to
include black hole rotation and our suite of numerical codes -- {\sc
  Titan/Noar/KY} -- allows to treat the black hole angular momentum as a
parameter.) 

For small distances to the black hole, light-bending effects must be important
\citep{miniutti2004} and the primary photons emerging from the flare source
are directed down to the disk without reaching a very far distance from the
flare. The light-bending is an effect of the curved space-time in the direct
vicinity of the black hole, and becomes weaker for flares at larger disk
radii. We also assume that the intrinsic radiation is collimated inside an
irradiation cone limited by $\theta = 60\deg$. A possible interpretation of
the irradiation cone can be given by assuming that the primary emission of the
flare is not isotropic. A simple argument for such an assumption has been
given by Nayakshin (2007) stating that magnetic flares occurring above the
accretion disk of AGN are as likely to be anisotropic as solar
flares. Nayakshin (2007) supports this argument with a simple consideration of
the radiative transfer inside the coronal flare source. It is likely that at
least for flares in the innermost region of the accretion disk both effects
play a role.

The results we present in this paper show that detailed radiative transfer
calculations for the reprocessed flare emission and its modifications due to
the relativistic effects in the vicinity of the black hole are computationally
feasible with current technology. They deliver a vast variety of results for
the time-dependent spectral energy distributions and for the flare
lightcurves. To apply them to a full extend to the observations of AGN, future
generations of X-ray observatories are necessary. The models we presented here
then allow to put important geometrical constraints on the flaring activity in
the accretion flow of supermassive black holes.

\begin{acknowledgements}
This work was supported in part by the grant 1P03D00829 of the Polish State
Committee for Scientific Research, by the Laboratoire Europe\' en Associ\' e
Astrophysique Pologne--France, and be the ESA-PECS project No.~98040. RG is
grateful to the Centre of Theoretical Astrophysics and to the
Hans-B\"ockler-Stiftung. MD and VK gratefully acknowledge support from the
Czech Science Foundation grants 205/05/P525 and 205/07/0052. 
\end{acknowledgements}

\end{document}